\makeatletter
\def\TODAY{\ifcase\month\or 01\or 02\or 03\or 04\or 05\or 
    06\or 07\or 08\or 09\or 10\or 11\or 12\fi/\number\day/\number\year}
\def\timenow{
  \@tempcnta=\time \divide\@tempcnta by 60 \number\@tempcnta:\multiply
  \@tempcnta by 60 \@tempcntb=\time \advance\@tempcntb by -\@tempcnta
  \ifnum\@tempcntb <10 0\number\@tempcntb\else\number\@tempcntb\fi}
\def\VERSION{\TODAY~at~\timenow:\ \ {\bf \jobname.tex}}
\makeatother
\documentclass[aps,pre,floats,superscriptaddress,twocolumn,amssymb,amsmath,showpacs,eqsecnum]{revtex4}
\usepackage{amssymb}
\usepackage{color}

\usepackage{epsf}
\usepackage{subfigure}
\usepackage{graphicx}
\usepackage{color}
\usepackage{epstopdf}
\usepackage[hypertex]{hyperref}
\begin{document}
\title{Boundary conditions and the critical  Casimir force on an Ising model film: exact results in one and two dimensions}

\author{Joseph Rudnick}

\affiliation{Department of Physics and Astronomy, UCLA, Box 951547 Los Angeles California 90095-1547}

\author{Roya Zandi}

\affiliation{Department of Physics and Astronomy, University of California, Riverside}

\author{Aviva Shackell}

\affiliation{Department of Physics and Astronomy, UCLA, Box 951547 Los Angeles California 90095-1547}

\author{Douglas Abraham}

\affiliation{Theoretical Physics, Department of Physics, University of Oxford,
1 Keble Road, Oxford OX1 3NP, United Kingdom}

\date{\today}

\begin{abstract}

Finite size effect in critical systems can be interpreted as universal Casimir forces.  Here we compare the Casimir force for free, periodic and antiperiodic boundary conditions in the exactly calculable case of the Ising model in one and two dimensions.  We employ a new procedure which allows us to calculate the Casimir force with periodic and antiperiodic boundary conditions analytically in a transparent manner. Quite interestingly, we find an attractive Casimir force for the case of periodic boundary condition but a repulsive Casimir force for the anti-periodic one.

\end{abstract}
\maketitle
\section{Introduction \label{sec:intro}}
Since the original work by Fisher and de Gennes \cite{fisherdegennes}, it has been well accepted that in the vicinity of a critical point the developing long range fluctuations give rise to a Casimir force in a system with a film-like geometry. As a consequence of finite size scaling \cite{fishbarb}, the generic form of the force is
\begin{equation}
\frac{F_{{ \rm Casimir}}}{k_BT_C} = \frac{1}{L^d} \vartheta\left(L/\xi\right)
\label{eq:gencas}
\end{equation}
where $T_C$ is the critical temperature, $L$ is the thickness of the film, $d$ is the bulk dimensionality of the system, $\xi$ is the correlation length which diverges at the critical point in the bulk, and $\vartheta$ is a universal function of its one argument. 

The notion of universality requires some qualification in the context of Casimir forces. Generally, for a system with a critical point, universality implies dependence only on the universality class to which that system belongs.  In the case of an Ising model, that universality class is characterized as $O(1)$, corresponding to the discrete symmetry that is broken in the ordered phase. Systems belonging to this class include uniaxial ferromagnets, simple liquid-vapor systems and systems exhibiting de-mixing \cite{binary}.

The universality class, however,  is not the only characteristic of the system that dictates the behavior of the function $\vartheta(x)$; one must also consider the boundary conditions at the two extremities of the film. This effect is dramatically exhibited in the results of measurements performed in Ref.  \cite{gc1,gc2}. Those experiments monitor the thickness of a film of $^4$He as the temperature passes through the lambda point at which bulk Helium acquires superfluid characteristics. It was found that the thickness of the film, which directly tracks the Casimir force appearing in the vicinity of this critical point, changes substantially when the temperature is close to and slightly below $T_{\lambda}$.   Quite interestingly, deep in the superfluid phase, the film partially recovers its thickness but still remains thinner than in the normal phase.

The thinning of the film well below the transition point  was successfully explained in Ref \cite{zandi2}, where it is shown that the confinement of bulk Goldstone modes combined with the effect of finite film thickness on surface fluctuations gives rise to a Casimir force that reduces the thickness of the film in the superfluid phase as compared to that of the normal film. However, the much more substantial thinning of the film right below the critical point remained unexplained for many years.

Earlier predictions of the Casimir force at the vicinity of $\lambda$ transition based on calculations emphasizing the correct universality class, $O(2)$ for the superfluid transition in $^4$He  \cite{KD,williams}, differ markedly from the experimentally determined force,  which is greater in magnitude by a factor of about thirty than those theoretical results. 

In Refs. \cite{zandi1,maciolek,hucht} it was shown that the discrepancy arises from the boundary conditions assumed in the previous calculations. In particular, prior field theoretical calculations \cite{krech,KD} as well as the simulations in \cite{KD} assume periodic boundary conditions, while physical reality  requires the superfluid order parameter to vanish immediately outside of the Helium film, thus mandating Dirichlet boundary conditions.  It has been known for a long time that the boundary conditions operative in a system play an important role in determining the behavior of Casimir forces \cite{danbook,miltonbook}. However, there is no {\em a priori} reason to assume that the strength of Casimir force with Dirichlet boundary conditions will be significantly larger than that for the case of periodic boundary conditions. In fact, the absence of this expectation undoubtedly hindered our understanding of the dip in Helium films for many years.

The strong effect of boundary conditions on the Casimir force in the particular instance of an $O(2)$ system in three dimensions leads one to ask if there are any general inferences one can arrive at regarding the interplay of boundary conditions, dimensionality, and universality class on the Casimir force acting on a system at and near the critical point. With an eye toward developing a greater depth of understanding of effects of boundary conditions and dimensionality, we have looked closely at one system for which exact results can be obtained: an Ising model in one and two dimensions.  With regard to the one-dimensional Ising model relevant results follow almost immediately from a simple transfer matrix analysis. In the case of the two dimensional Ising model, we have been able to perform an analysis of that model that yields the critical Casimir force for four different boundary conditions: periodic, antiperiodic, free (the appropriate surrogate for Dirichlet boundary conditions in the Ising model) and fixed-spin. The results of this analysis are consistent with those we have extracted from various expressions already in the literature.  

A key conclusion that we draw is that boundary conditions alone do not determine the relative amplitudes of critical Casimir forces. Indeed, we find that as dimensionality is reduced, periodic (and antiperiodic) boundary conditions give rise to a critical Casimir force that is increasingly stronger than the force resulting from free boundary conditions. In fact, the critical Casimir force for a one dimensional Ising model with free boundary conditions is effectively non-existent.

The paper is organized as follows.
In section \ref{onedimension} we calculate the Casimir force for the one-dimensional Ising model as a prelude to the discussion of the two dimensional version. We obtain the Casimir forces with free, fixed, periodic and anti-periodic boundary conditions. We observe significant differences in the forces in these four cases.   In section \ref{twodimension}, we present our results for the two dimensional Ising model and compare them with the one dimensional results. The forces are derived in two different ways. First, we show how they follow from expressions already in the literature. We also demonstrate how they arise from an unified analysis of the two dimensional Ising model making use of an approach along the lines of the method developed by Schultz, Mattis and Lieb \cite{SML}. In section \ref{conclusion}, we discuss our findings, their implications, and summarize our conclusions. For clarity of exposition, details of our calculations are relegated to appendices. 

\section{one dimension}
\label{onedimension}

\subsection{periodic boundary conditions}

The partition function of the one dimensional Ising model is well known to be derivable from the two-by-two transfer matrix of that model. The Ising model possesses a global $Z_2$ symmetry and is described by the following Hamiltonian
\begin{equation}
{\cal H} =-J \sum_{i,j}s_is_j,
\label{energy}
\end{equation}
where $J>0$ is the spin-spin coupling constant, and  the sum is over nearest neighbor bonds $i,j$ on the lattice.  The spin variables $s_{i}$ can assume the values $\pm1$. Given the transfer matrix
\begin{equation}
{ \bf T}_1 = \left(\begin{array}{ll} e^{\beta J} & e^{- \beta J} \\ e^{- \beta J} & e^{\beta J} \end{array} \right)
\label{eq:T1}
\end{equation}
where $\beta=1/k_BT$, a spin state at the left boundary of the $N$-spin array of the form $\langle l|$ and a spin state $|r \rangle$ at the right boundary of that array, the partition function is equal to
\begin{equation}
\mathcal{Z } = \langle l| { \bf T}_1^N |r \rangle
\label{eq:genpart1}
\end{equation}
The quantity on the right hand side of (\ref{eq:genpart1}) is most conveniently evaluated making use of the eigenvector decomposition of ${ \bf T}_1$
\begin{eqnarray}
{ \bf T}_1 & = & |e \rangle 2 \cosh (\beta J) \langle e| + |o \rangle 2 \sinh (\beta J) \langle o|
\label{eq:T2decomp}
\end{eqnarray}
where 
\begin{eqnarray}
|e \rangle & = &\frac{1}{\sqrt{2}} \left( \begin{array}{rr} 1 \\ 1 \end{array} \right) \label{eq:eveneig} \\
|o \rangle & = & \frac{1}{\sqrt{2}} \left( \begin{array}{rr} 1 \\ -1 \end{array} \right) \label{eq:oddeig}
\end{eqnarray}

When periodic boundary conditions hold, we calculate the partition function by evaluating the trace of ${ \bf T}_1^N$ over a complete set of bounding spin states. Given (\ref{eq:eveneig}) and (\ref{eq:oddeig}) we have immediately
\begin{equation}
\mathcal{Z}_{\rm periodic}(N) = 2^{N} \left[  \cosh^N(\beta J) + \sinh^N(\beta J)\right],
\label{eq:part1}
\end{equation}
Therefore, the free energy of this system is
\begin{eqnarray}
\lefteqn{\mathcal{F}_{\rm periodic}(N)} \nonumber \\ & = &  -N k_BT \ln \left( 2\cosh( \beta J) \right)- k_BT\ln \left[1+\tanh(\beta J)^N\right] \nonumber \\
\label{eq:free1}
\end{eqnarray}
In the thermodynamic limit the second term of (\ref{eq:free1}) becomes negligible, so the bulk free energy per unit length (alternatively per spin) is 
 \begin{equation}
f_B =-k_BT \ln 2 - k_BT \ln \cosh (\beta J)
\label{eq:free2}
\end{equation}

The correlation function, $C(l_1,l_2)$, and the associated correlation length, $\xi$,   in the one-dimensional Ising model are defined as follows \cite{baxter}
\begin{eqnarray}
C(l_1,l_2) & = &  \tanh(\beta J)^{l_2-l_1} \nonumber \\
& \equiv & \exp \left[ -(l_2-l_1)/ \xi\right]
\label{eq:corr2}
\end{eqnarray}
which leads to 
\begin{equation}
 \xi= -1/(\ln \tanh \beta J)
 \label{eq:correlationlength}
\end{equation}
Taking the derivative of the free energy (\ref{eq:free1}) with respect to $N$ and subtracting the bulk free energy per unit length, we are left with
\begin{eqnarray}
-\frac{\partial \mathcal{F}(N)}{\partial N} + f_B &=&- \frac{k_BT}{\xi} \frac{e^{-N/\xi}}{1+e^{N/ \xi}} \nonumber \\
& = & -\frac{k_BT}{N} \left\{\frac{N}{\xi} \frac{e^{-N/\xi}}{1+e^{N/ \xi}}\right\}
\nonumber  \\ & = & F^{\rm Cas}_{1d \ {\rm per}}
\label{eq:scform1}
\end{eqnarray}
The Casimir force has precisely has the desired form (\ref{eq:gencas}).  That is, for the one-dimensional Ising model
\begin{equation}
\vartheta_{1d \ {\rm per}}(x) = -x\frac{e^{-x}}{1+e^{-x}}
\label{eq:onedper}
\end{equation}

Note that the factor $k_BT$ plays an essential role in our result for the Casimir force. To the extent that it can be said to exist, the critical point of the one-dimensional Ising model is at $T=0$. Unless we normalize the critical Casimir force in terms of this temperature-dependent factor, we are forced to conclude that it vanishes.

\subsection{Antiperiodic boundary conditions}

The second boundary condition we study here is antiperiodic. We enforce this boundary condition by introducing an array of antiferromagnetic bonds in the system.  In this case, there are $N-1$ two-by-two transfer matrices of the form (\ref{eq:T1}) 
and one transfer matrix of the form
\begin{equation}
{ \bf T}_2 = \left(\begin{array}{ll} e^{-\beta J} & e^{ \beta J} \\ e^{ \beta J} & e^{-\beta J} \end{array} \right) \label{eq:T2}
\end{equation}
The eigenvector decomposition of the transfer matrix ${ \bf T}_2$ is
\begin{equation}
{ \bf T}_2 = |e \rangle2  \cosh (\beta J)\langle  e | - |o \rangle2   \sinh (\beta J)\langle o |
\label{eq:T2a}
\end{equation}

\begin{figure}[htbp]
\begin{center}
\includegraphics[width=3in]{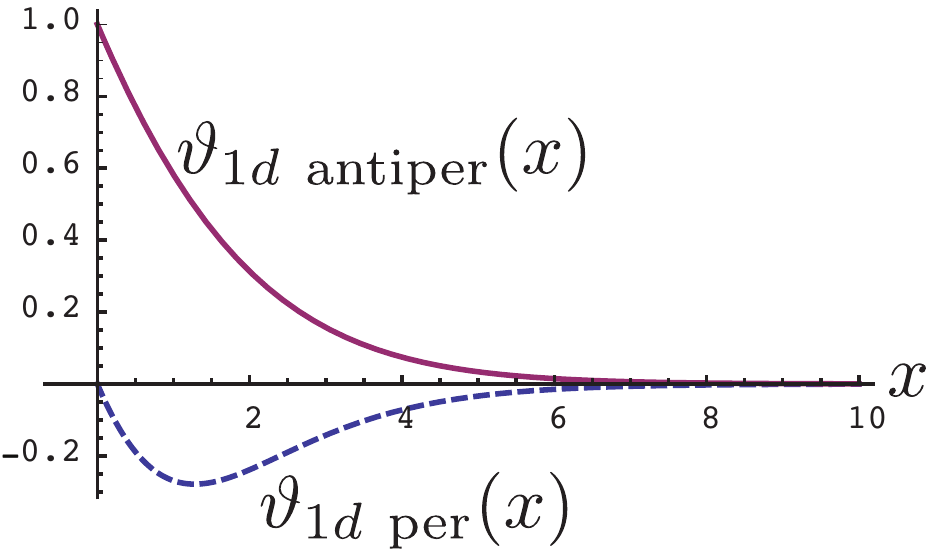}
\caption{The universal Casimir force function $\vartheta_{1d \ {\rm per}}(x)$, shown as a dashed curve and the function $\vartheta_{1d \ {\rm antiper}}(x)$, shown as a solid curve. }
\label{fig:apGplot}
\end{center}
\end{figure}

The partition function of this system is the trace of the quantity ${ \bf T}_1^{N-1} \cdot { \bf T}_2$. Making use of (\ref{eq:eveneig}) and (\ref{eq:oddeig}), we have 
\begin{eqnarray}
\mathcal{Z}_{\rm antiper} & = & \mathop{\rm Tr}\left( { \bf T}_1^{N-1} \cdot { \bf T}_2 \right) \nonumber \\
& = & (2 \cosh \beta J)^N - ( 2\sinh  \beta J)^N
\label{eq:partap1}
\end{eqnarray}
This means that the free energy is given by
\begin{eqnarray}
\mathcal{F} & = & - k_BT \ln \mathcal{Z} \nonumber \\
& = & - N k_BT  \ln \left(2 \cosh \beta J \right) - k_BT \ln \left[ 1- (\tanh \beta J)^N\right] \nonumber \\
& = & N f_B -k_bT \ln \left[ 1-e^{-N/ \xi}\right]
\label{eq:freeap1}
\end{eqnarray}
The quantity $f_B$ in the last line of (\ref{eq:freeap1}) is the ``bulk'' free energy of the one-dimensional Ising model. The quantity $\xi= -1/(\ln \tanh \beta J)$ is the correlation length of that model. We then take the negative derivative of the free energy with respect to $L$, add in $f_B$, and end up with
\begin{eqnarray}
F^{\rm Cas}_{1d \ {\rm antiper}} & = & - \frac{\partial \mathcal{F}}{\partial N} +f_B \nonumber \\ & = & \frac{k_B T }{\xi} \frac{e^{-N/\xi}}{1-e^{-N/ \xi}} \nonumber \\
& = & \frac{k_BT}{N} \frac{N}{\xi}\frac{e^{-N/\xi}}{1-e^{-N/ \xi}}  \nonumber \\
&\equiv & \frac{k_BT}{N} \vartheta_{1d \ {\rm antiper}}(N/ \xi)
\label{eq:scalecasap1}
\end{eqnarray}
Fiture \ref{fig:apGplot} displays universal functions $\vartheta_{1d \ {\rm per}}(x)$ and $\vartheta_{1d \ {\rm antiper}}(x)$.
Two distinct features of this scaling function distinguish it from the case of periodic boundary conditions; it is positive, indicating a repulsive Casimir force, and it does not go to zero at the zero temperature critical point.

\subsection{Free and Fixed Boundary Conditions} 

In the case of free boundary conditions, the partition function of the $N$-layer system is given by
\begin{eqnarray}
\mathcal{Z}_{\rm free}  & = &  2  \langle e | { \bf T}_1^N | e \rangle \nonumber \\
& = & 2 ( 2 \cosh (\beta J))^N
\label{eq:freebc1}
\end{eqnarray}
Taking the derivative with respect to $N$ of the log of the above result and subtracting the bulk contribution to that derivative, we find perfect cancellation. There is no residual force. 

When the spins are fixed at the boundary, the state between which the $N^{\rm th}$ power of the transfer matrix is sandwiched is
\begin{equation}
|f \rangle = \left( \begin{array}{l} 1 \\ 0 \end{array} \right)
\label{eq:fixed1}
\end{equation}
and the partition function is given by
\begin{eqnarray}
\lefteqn{\mathcal{Z}_{\rm fixed}} \nonumber \\  & = & \langle f | { \bf T}_1^N | f \rangle \nonumber \\
& = & \frac{1}{2} \left(( 2 \cosh (\beta J))^N + ( 2 \sinh (\beta J))^N \right)
\label{eq:fixed2}
\end{eqnarray}
Given the near identity between the final line of (\ref{eq:fixed2}) and what was obtained for the partition function of the system with periodic boundary conditions in (\ref{eq:part1}), we find immediately that the critical Casimir force induced by fixed boundary conditions matches the corresponding force in the case of periodic boundary conditions.

\section{Two dimensions}
\label{twodimension}
In this section, we consider an Ising model on a two dimensional square lattice consisting of $M$ rows and $N$ columns.  The Hamiltonian of this model is the same as given in Eq. (\ref{energy}) but the sum is over a two dimensional lattice.  We will begin by focusing on periodic and antiperiodic boundary conditions, first demonstrating how the critical Casimir force follows from expressions for the partition function already in the literature. Then, we perform an analysis of the two dimensional Ising model based on the fermion variable approach introduced by Schultz, Mattis and Lieb that leads to precisely the same results. Finally, we make use of this same approach to rederive existing expressions for the critical Casimir force in the case of free and fixed boundary conditions. As has already been noted, the Casimir forces in the two dimensional Ising model for free and fixed-spin boundary conditions  are connected by a duality relationship.

\subsection{Periodic boundary conditions}
\label{sec:periodic2d}

Our starting point in the calculation of periodic boundary conditions is the formula utilized by Ferdinand and Fisher \cite{ferdfish} for the partition function of a $M\times N$ square Ising lattice wrapped on a torus.
\begin{eqnarray}
\lefteqn{\exp \left[\sum_{r=1}^{N-1} \ln \cosh \frac{1}{2} M \gamma_{2r+1} \right] +\exp \left[\sum_{r=1}^{N-1} \ln \sinh \frac{1}{2} M \gamma_{2r+1} \right]} \nonumber \\ && + \exp \left[\sum_{r=1}^{N-1} \ln \cosh \frac{1}{2} M \gamma_{2r} \right] + \exp \left[\sum_{r=1}^{N-1} \ln \sinh \frac{1}{2} M \gamma_{2r} \right]  \nonumber\\
\label{eq:part2}
\end{eqnarray}
In (\ref{eq:part2})
\begin{equation}
\gamma_l = \ln \left[  c_l + \sqrt{c_l^2-1}\right]
\label{eq:gammal}
\end{equation}
where
\begin{eqnarray}
c_l  & = & \cosh 2 K \coth 2 K - \cos l \pi/n \nonumber \\
 & = &   \cosh 2K \coth 2K - \frac{1}{2}(z_l+z_l^{-1}) \nonumber \\
& \equiv & c(z_l)
\label{eq:cont2}
\end{eqnarray}
with
\begin{equation}
K=\frac{J}{k_BT}
\label{eq:Kdef}
\end{equation}
and
\begin{equation}
z_l=e^{i l \pi/n}
\label{eq:cont3}
\end{equation}

Given that we are interested in the limit $N \rightarrow \infty$, corresponding to the thermodynamic limit for the width of the of the two dimensional strip with finite thickness $M$, we can replace the sums in the exponents of (\ref{eq:part2}) by integrals. This means that the partition function in question reduces to 
\begin{eqnarray}
2 \exp \left[\frac{N}{2 \pi} \int_0^{2 \pi} \ln \sinh \left(\frac{M}{2} \gamma(e^{i \phi}) \right) d \phi \right] +\nonumber \\
2 \exp \left[\frac{N}{2 \pi} \int_0^{2 \pi} \ln \cosh \left(\frac{M}{2} \gamma(e^{i \phi}) \right) d \phi \right]
\label{eq:cpart1}
\end{eqnarray}
which can be rewritten as follows:
\begin{widetext}
\begin{eqnarray}
\lefteqn{2 \exp \left[\frac{N}{2 \pi} \int_0^{2 \pi} \ln \cosh \left(\frac{M}{2} \gamma(e^{i \phi}) \right) d \phi \right] } \nonumber \\ && \times \left(1+ \exp \left[\frac{N}{2 \pi} \left(\int_0^{2 \pi} \ln \sinh \left(\frac{M}{2} \gamma(e^{i \phi}) \right) d \phi- \int_0^{2 \pi} \ln \cosh \left(\frac{M}{2} \gamma(e^{i \phi}) \right) d \phi\right) \right]\right) \nonumber \\ & = & 2 \exp \left[\frac{N}{2 \pi} \int_0^{2 \pi} \ln \cosh \left(\frac{M}{2} \gamma(e^{i \phi}) \right) d \phi \right]   \left\{1+  \exp \left[\frac{N}{2 \pi} \int_0^{2 \pi} \ln \tanh \left(\frac{M}{2} \gamma(e^{i \phi}) \right) d \phi  \right]\right\}
\label{eq:ans1}
\end{eqnarray}
\end{widetext}
Because $\tanh x <1$ for any positive $x$, the function $\ln \tanh x$ will be negative. Thus the integral on the last line of (\ref{eq:ans1}) produces a negative number, which is multiplied by $N$, the transverse extent of the strip, the analog of the area of a three dimensional film. We are interested in the limit $n\rightarrow \infty$, so the second term in curly brackets on the last line of (\ref{eq:ans1}) will be negligible compared to unity, and the sinh contribution to the partition function in (\ref{eq:cpart1}) can be discarded without error.

Taking the log and then the derivative with respect to $m$, we end up with the following integral
\begin{equation}
-\frac{\partial \mathcal{F}}{\partial M} = \frac{k_BT}{\pi} \int_0^{2 \pi} \gamma(e^{i \phi}) \tanh M \gamma(e^{i \phi}) d \phi
\label{eq:cforce1}
\end{equation}
Making use of the fact that $\lim_{x \rightarrow  \infty }\tanh x  = 1$, we calculate the force for the bulk system which we subtract from (\ref{eq:cforce1}) to find the following Casimir force
\begin{eqnarray}
\lefteqn{\frac{\vartheta_{2d {\rm per}}}{k_BT}} \nonumber \\ & = & 
\frac{1}{\pi} \int_0^{2 \pi} \gamma(e^{i \phi}) \tanh M \gamma(e^{i \phi}) d \phi - \frac{1}{\pi} \int_0^{2 \pi} \gamma(e^{i \phi}) d \phi \nonumber \\ 
\label{eq:cforce2}
\end{eqnarray}

\subsection{The scaling limit} \label{page:sstart}
To obtain the scaling form of the Casimir force, we start with looking at the behavior of the function $\gamma(e^{i \phi})$ in the vicinity of the critical point at which
\begin{equation}
\cosh 2K \coth 2K=2.
\label{eq:critK}
\end{equation}
The solution to the equation (\ref{eq:critK}) is given by
\begin{eqnarray}
\sinh ^2K_C &=&1     \nonumber \\    K_C&=&\frac{1}{2}\ln(1+\sqrt(2))=0.44068...
\label{eq:KC}
\end{eqnarray}

\noindent as given in Ref.\cite{ferdfish}.  We now introduce the reduced temperature, $\tau$:
\begin{eqnarray}
\tau
& = & -\mathop {\rm sgn} \left(  \frac{1}{2}\cosh 2K \coth 2K -1 \right)  \nonumber \\ && \times \sqrt{ \frac{1}{2}\cosh 2K \coth 2K -1}
\label{eq:tau}
\end{eqnarray}
Then for small $\tau$ and $\phi$, $c(e^{i \phi}) = 1+2 \tau^2  + \frac{\phi^2}{2}$, and for the corresponding expansion of the function $\gamma(c(e^{i \phi}))$, we are left with
\begin{eqnarray}
\gamma(c(e^{i \phi})) & \rightarrow & \ln\left[ 1 + 2\sqrt{ \tau^2 + \phi^2/4} \right] \nonumber \\
& \rightarrow & 2 \sqrt{ \tau^2 + \phi^2/4}
\label{eq:gammaexpand2}
\end{eqnarray}
If we set $\phi=2 \omega$ and replace $\omega$ by $\Omega/M$ and $\tau$ by $x/M$, we find the following expression for the Casimir force t
\begin{equation}
\frac{1}{\pi M^2} \int_{-\infty}^{\infty} d \Omega \sqrt{x^2 + \Omega^2}\left[ \tanh \sqrt{x^2+ \Omega^2} - 1 \right]
\label{eq:ccas2}
\end{equation}
The universal function $\vartheta(x)$ that emerges from (\ref{eq:ccas2}) is
\begin{equation}
\vartheta_{2d {\rm per}}(x) = \frac{1}{ \pi }\int_{- \infty}^{\infty} d \Omega \sqrt{x^2 +\Omega^2} \left[ \tanh \sqrt{x^2+ \Omega^2} -1 \right] 
\label{eq:ccas3}
\end{equation}
This result is displayed as the short dashed curve in \mbox{Fig. \ref{fig:compare3}}.

\subsection{The two dimensional Ising model with antiperiodic boundary conditions} \label{sec:2danti}

The exact partition function for a two dimensional Ising model on  $M \times N$ square lattice with anti-periodic-antiperiodic boundary conditions is
\begin{eqnarray}
\lefteqn{-\exp \left[\sum_{r=1}^{N-1} \ln \cosh \frac{1}{2} M \gamma_{2r+1} \right] +\exp \left[\sum_{r=1}^{N-1} \ln \sinh \frac{1}{2} M \gamma_{2r+1} \right]} \nonumber \\ && + \exp \left[\sum_{r=1}^{N-1} \ln \cosh \frac{1}{2} M \gamma_{2r} \right] - \exp \left[\sum_{r=1}^{N-1} \ln \sinh \frac{1}{2} M \gamma_{2r} \right] \nonumber \\
\label{eq:part3}
\end{eqnarray}
The derivation of the above partition function is given in Appendix \ref{app:wuhu} and is based on the expression calculated by Wu and Hu in Ref. \cite{wuandhu}. Note that the last term in (\ref{eq:part3}) changes sign at the critical temperature; see Appendix \ref{app:wuhu} for more detail. As previously, quantity $M$ is the width of the strip, while $N$ is its extent. Again, we will take the limit $N \rightarrow \infty$ before allowing $M$ to become large. 

The difference between periodic and antiperiodic boundary conditions lies in the sign relationships between the terms in Eqs. (\ref{eq:part2}) and (\ref{eq:part3}). Consider the two terms with the log of the hyperbolic cosine in Eq. (\ref{eq:part3}). They appear with opposite signs, and as we will see, this leads to near-cancellation of their contributions to the partition function. On the other hand, the other two terms in (\ref{eq:part3}) may or may not add up. In order to assess the full partition function, we need to look at what remains after all near cancellations are taken into account. 

Appendix \ref{sec:details} contains an analysis of the contributions of the various terms in (\ref{eq:part3}) to the partition function. The end-result is that the hyperbolic sine terms now dominate. We take a derivative with respect to $M$, the film thickness, to obtain the Casimir force. Performing the same scaling analysis as in Section \ref{sec:periodic2d}, we find for the Casimir force in the case of antiperiodic boundary conditions.
\begin{equation}
\frac{1}{\pi M^2} \int_{-\infty}^{\infty} d \Omega \sqrt{x^2+ \Omega^2} \left[ \coth \sqrt{t^2 + \Omega^2} -1 \right] 
\label{eq:cas2dap}
\end{equation}
For comparison, see (\ref{eq:ccas2}) for the Casimir force in the case of periodic boundary conditions. Figure \ref{fig:compare3} shows the Casimir forces for periodic, antiperiodic and free boundary conditions. Note that when antiperiodic boundary conditions apply, the critical Casimir force is repulsive, leading to film thickening, as was the case for the one dimensional Ising model. 

\subsection{Free boundary conditions}

To obtain the Casimir force for a two dimensional Ising model in the case of free boundary conditions, we use the expressions used in Li, {\em et. al.} \cite{li}.  After some manipulations, we find the asymptotic scaling form of  the critical portion of the free energy per unit area is
\begin{eqnarray}
\lefteqn{F_{\rm singular}} \nonumber \\ & = &  -\frac{1}{\pi} \int_0^1 \frac{d \omega}{\sqrt{1- \omega^2}}\Bigg\{2 \ln (1-Z_S)^2  \nonumber \\ && + \ln \Bigg[ \frac{1}{2} \left( 1+\frac{\tau}{\sqrt{\tau^2 + \omega^2}}\right) \left(1+ 2 \sqrt{\tau^2 + \omega^2} \right) ^N \nonumber \\ && + \frac{1}{2} \left( 1-\frac{\tau}{\sqrt{\tau^2 + \omega^2}}\right) \left(1- 2 \sqrt{\tau^2 + \omega^2} \right) ^N\Bigg] \Bigg\} \nonumber \\
\label{eq:2bc9}
\end{eqnarray}

\begin{figure}[htbp]
\begin{center}
\includegraphics[width=3in]{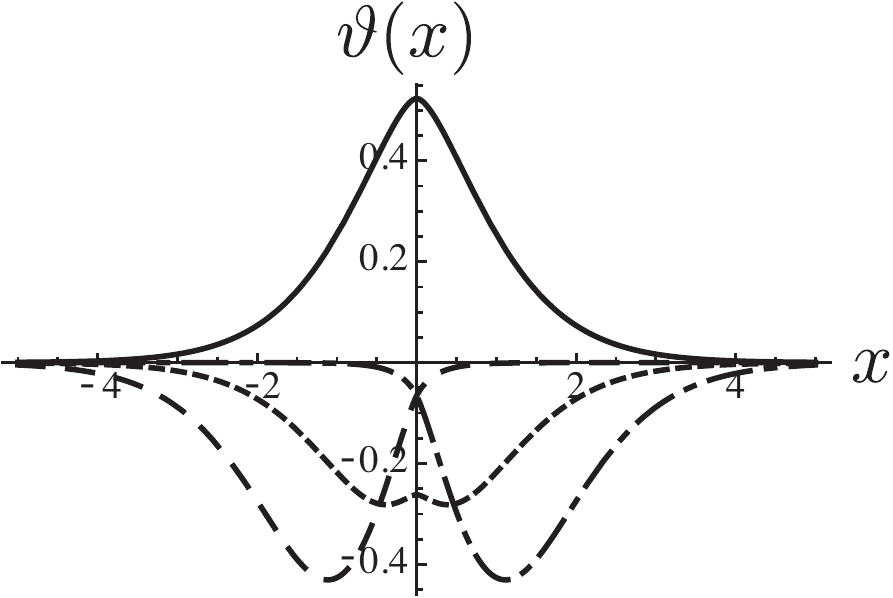}
\caption{The scaling contribution in the two dimensional Ising model to the Casimir force for antiperiodic boundary conditions (solid curve), periodic boundary conditions (short dashed curve) and free boundary conditions (long dashed curve) and fixed spin boundary conditions (alternatively long and short dashed curve).}
\label{fig:compare3}
\end{center}
\end{figure}

Here $Z_S=\tanh(\beta J_S)$  with $J_S$ the in-layer coupling on the surface and  $\tau$ is the reduced temperature.    In a change of notation, we now denote the thickness of the film by $N$. For simplicity, we assume that  bulk couplings in both $x$ and $y$ directions are equal.  Note that the reduced temperature in the above equation is equal to $1/N$ of  the reduced  temperature defined in Li {\em et. al.} \cite{li}.  To find the Casimir force, we take the negative of the derivative of the free energy above with respect to $N$. Since $N$ appears only in the exponent, we need to consider the following derivative
\begin{eqnarray}
\lefteqn{\frac{\partial}{\partial N} \left(1+ 2 \sqrt{\tau^2 + \omega^2} \right) ^N=}    \nonumber \\&& \left[\ln \left(1+ 2 \sqrt{\tau^2 + \omega^2} \right)  \right] \left(1+ 2 \sqrt{\tau^2 + \omega^2} \right) ^N
\label{eq:2bc10}
\end{eqnarray}
which helps us introduce the relevant scaling variables. If we replace the reduced temperature, $\tau$, by $x/N$ and the integration variable, $\omega$, by $\Omega/N$, the right hand side of (\ref{eq:2bc10}) in the limit of large $N$ becomes
\begin{equation}
\frac{2}{N}\sqrt{x^2+ \Omega^2}e^{2 \sqrt{x^2+ \Omega^2}}
\label{eq:2bc11}
\end{equation}
and thus the negative derivative of the free energy (Eq. \ref{eq:2bc9}) is
\begin{eqnarray}
\lefteqn{-\frac{\partial F_{ \rm singular}}{\partial N} = \frac{2}{\pi N^2} \int_0 d \Omega \sqrt{x^2 + \Omega^2} \times} \nonumber \\&& \frac{\left(1+\frac{x}{\sqrt{x^2 + \Omega^2}} \right)e^{2 \sqrt{x^2+ \Omega^2}}-\left(1-\frac{x}{\sqrt{x^2 + \Omega^2}} \right)e^{-2 \sqrt{x^2+ \Omega^2}}}{\left(1+\frac{x}{\sqrt{x^2 + \Omega^2}} \right)e^{2 \sqrt{x^2+ \Omega^2}}+\left(1-\frac{x}{\sqrt{x^2 + \Omega^2}} \right)e^{-2 \sqrt{x^2+ \Omega^2}}} \nonumber\\
\label{eq:2bc12}
\end{eqnarray}
Now we need to subtract from the above expression the "background" casimir forces which is obtained when the thickness of the film is infinite.  As $N \rightarrow \infty$, the free energy given in (\ref{eq:2bc9}) becomes dominated by the term going as $(1+2 \tau/\sqrt{\tau^2+ \omega^2})^N$, and thus the corresponding Casimir force per unit area is
\begin{equation}
\frac{1}{\pi} \int_0^1 \frac{d \omega}{\sqrt{1- \omega^2}} \ln \left( 1+ 2 \sqrt{\tau^2+ \omega^2}\right)
\label{eq:2bc13}
\end{equation}
Inserting the scaling variables $x/N=\tau$ and $\omega=\Omega/N$ as before, we obtain
\begin{equation}
\frac{2}{ \pi N^2} \int_0 d \Omega \sqrt{x^2+ \Omega^2}
\label{eq:2bc14}
\end{equation}
Subtracting this from (\ref{eq:2bc12}), we are now allowed to extend the upper limit of integration to $\infty$, and thus we end up with the final expression for the Casmir force per unit area
\begin{widetext}
\begin{eqnarray}
\mathcal{F}  =  \frac{1}{\pi N^2} \int_{-\infty}^{\infty} d \Omega \sqrt{x^2 + \Omega^2}  \left [ \frac{\left(1+\frac{x}{\sqrt{x^2 + \Omega^2}} \right)e^{2 \sqrt{x^2+ \Omega^2}}-\left(1-\frac{x}{\sqrt{x^2 + \Omega^2}} \right)e^{-2 \sqrt{x^2+ \Omega^2}}}{\left(1+\frac{x}{\sqrt{x^2 + \Omega^2}} \right)e^{2 \sqrt{x^2+ \Omega^2}}+\left(1-\frac{x}{\sqrt{x^2 + \Omega^2}} \right)e^{-2 \sqrt{x^2+ \Omega^2}}} -1 \right] 
\label{eq:2bc15}
\end{eqnarray}
\end{widetext}
which is plotted in Fig. \ref{fig:compare3} (long dashed line) and as expected is negative giving rise to an attractive Casimir force. It is interesting to note that the strength of the Casimir force in the case of free boundary condition is zero in one dimension but in two dimensions is comparable in  amplitude to the Casimir force with periodic boundary conditions.  
We also anticipate the results of the next section by plotting the Casimir force for fixed spin boundary conditions, previously obtained by Evans and Stecki \cite{es}, which is related to the Casimir force for free boundary conditions by duality. 

\section{Alternative approach to the two dimensional Ising model}

It is also possible to obtain all the expressions above by a set of direct calculations of the partition function of the two dimensional Ising model in zero magnetic field. We make use of the method introduced by Schultz, Mattis and Lieb \cite{SML}, which will hereinafter be referred to as the LSM method. The version of this model we will address is wrapped on a cylinder with free boundary conditions on the ends. Circumferentially, we will allow for either periodic or antiperiodic boundary conditions. In this way, we encompass all the boundary conditions that were considered above. Additionally, we will demonstrate by a duality transformation that the critical Casimir force for free boundary conditions is simply related to the Casimir force for fixed boundary conditions obtained previously by Evans and Stecki \cite{es}. 

Denoting the partition function for free boundary conditions by $Z_f$, our first step is to write $Z_f$  in terms of a transfer matrix along the axis of a cylinder with circumference $M$ and length $N$. 
In the representation with two dimensional Pauli spin matrix $\sigma^x_j$, diagonal, $j=1, \dots, M$, the transfer operator within a row is given by
\begin{equation}
{ \bf V}_2 = \exp \left[K_2 \sum_{j=1}^M \sigma_j^x \sigma_{j+1}^x\right]
\label{eq:a1}
\end{equation}
and between rows by
\begin{equation}
{ \bf V}_1=  \exp \left[ -K_1^* \sum_{j=1}^M \sigma_j^z \right]
\label{eq:a2}
\end{equation}
In (\ref{eq:a1}) and (\ref{eq:a2}), $K_n$ with $  n=1,2$ are, respectively, the vertical and horizontal couplings, both positive for the ferromagnetic case, and $K_n^*$ is the \emph{dual} of $K_n$, i.e.
\begin{equation}
\tanh K_n^* = \exp \left[ -2K_n\right]
\label{eq:a3}
\end{equation}
Writing ${ \bf V}_1$ in the form (\ref{eq:a2}) requires that we accompany it by a factor $( 2 \sinh(2K_1))^{M/2}$ whenever it appears in an expression generating the partition function. In this article we make use of  the following representations of the Pauli spin matrices $\sigma_x$ and $\sigma_z$
\begin{eqnarray}
\sigma^x & = & \left( \begin{array}{rr} 1 & 0 \\ 0 & -1 \end{array}\right) \label{eq:sigmax} \\
\sigma^y & = & \left( \begin{array}{rr} 0 & -i \\ i & 0 \end{array}\right) \label{eq:sigmay} \\
\sigma^z & = & \left( \begin{array}{rr} 0 & -1 \\ -1 & 0 \end{array}\right) \label{eq:sigmaz} \\
\end{eqnarray}

\begin{figure}[htbp]
\begin{center}
\includegraphics[width=3in]{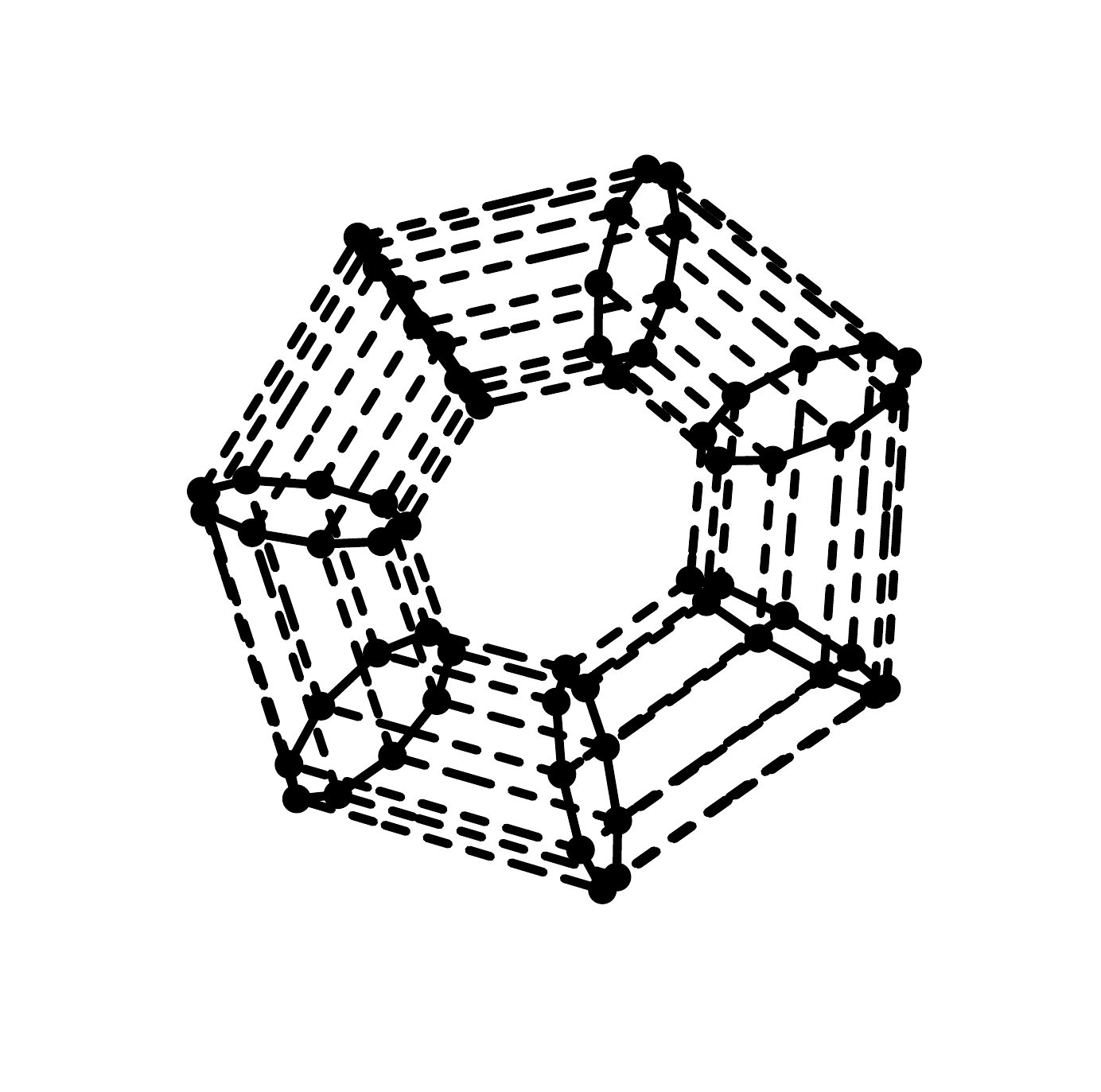}
\caption{The $M \times N$ lattice with $M=9$ and $N=7$, shown as a torus to indicate the cyclical nature of the vertical bond arrangement. The vertical bonds between the seven rows are shown dashed. }
\label{fig:torus1}
\end{center}
\end{figure}

The cylinder is assumed to couple to itself, so the actual depiction of the system ought to be as a torus, as shown in Fig. \ref{fig:torus1}.
Free boundary conditions are enforced by changing the strength of one array of vertical bonds to $K_0$ and then by taking the limit $K_0 \rightarrow 0$. The partition function $Z_f$, is then given by 
\begin{eqnarray}
\lefteqn{Z_f} \nonumber \\ &=& ( 2 \sinh (2K_1))^{M(N-1)/2} \lim_{K_0 \rightarrow 0}(2 \sinh 2K_0)^{M/2}  \nonumber \\ && \times \mathop{\rm Tr} \left(({ \bf V}_2 { \bf V}_1)^{N-1}{ \bf V}_2 \exp \left[ -K_0^* \sum_{m=1}^M \sigma_m^z\right] \right)
\label{eq:a4}
\end{eqnarray}
Given the fact that $K_0^* \rightarrow \infty$ as $K_0 \rightarrow 0$, we find
\begin{equation}
\lim_{K_0 \rightarrow 0}  \left(e^{-MK_0^*} \exp \left[-K_0^*\sum_{m=1}^M \sigma_m^z \right] \right) = |0 \rangle \langle 0 |
\label{eq:a5}
\end{equation}
where $|0\rangle$ is defined by
\begin{equation}
\sigma^z_j|0\rangle = - |0 \rangle, \ \ j=1, \ldots, M
\label{eq:a6}
\end{equation}
Furthermore, one can readily verify that as $K_0 \rightarrow 0$, $(2 \sinh 2K_0) \exp (2K_0^*) \rightarrow 4$. Cyclicity of the trace then reduces (\ref{eq:a4}) to
\begin{eqnarray}
Z_f
& = & (2 \sinh 2K_1)^{M(N-1)/2} 2^M \nonumber \\ &&  \times  \langle 0 | ({ \bf V}_2 { \bf V}_1)^{N-1}{ \bf V_2} |0 \rangle
\label{eq:ap1}
\end{eqnarray}
The final step is to insert ${ \bf V}_1^{-1/2}{ \bf V}_1^{1/2} = { \bf 1}$ at the beginning and end of the product of operators in (\ref{eq:ap1}) and then use $ { \bf V}_1^{1/2} |0 \rangle = \exp(-MK_1^*/2)|0 \rangle$. This yields
\begin{eqnarray}
Z_f & = & ( 2 \sinh 2K_1)^{M(N-1)/2} 2^M e^{-MK_1^*} \nonumber \\ &&  \times \langle 0| ({ \bf V}^{\prime})^{N} |0 \rangle
\label{eq:ap2}
\end{eqnarray}
where the operator ${ \bf V}^{\prime}$ is given by
\begin{equation}
{ \bf V}^{\prime} = { \bf V}_1^{1/2} { \bf V}_2 { \bf V}_1^{1/2}
\label{eq:ap3}
\end{equation}

Given that the operator ${ \bf V}^{\prime}$ is self adjoint, a natural step is to make use of its eigenvalue decomposition as determined by Schultz, Mattis and Lieb \cite{SML}. Before doing this, we will show how the strip with ``hard'' walls admits of a similar representation. We consider the same toroidal arrangement. This time, we alter the interactions between the spins in one of the $M$ spin circles. That is, we replace the interaction strengths in one of the ${ \bf V}_2$ terms by $K_0$ where we now take $K_0$ to infinity. In addition, we add a term to the Boltzmann constant that keeps the partition function finite in the $K_0 \rightarrow \infty$ limit.  Then, (\ref{eq:a4}) is replaced by
\begin{eqnarray}
\lefteqn{Z_h} \nonumber \\ & = & (2 \sinh 2K_1)^{MN/2} \lim_{K_0 \rightarrow \infty} \mathop{\rm Tr} \Bigg(\left({ \bf V}_1 { \bf V}_2 \right)^{N-1} { \bf V}_1 \nonumber \\ && \exp \left[ K_0 \sum_{m=1}^M \left(\sigma^x_m \sigma^x_{m+1}-1 \right)\right] \Bigg )
\label{eq:ap4}
\end{eqnarray}
Given that
\begin{equation}
\lim_{K_0 \rightarrow \infty} \exp \left[ K_0 \sum_{m=1}^M \left(\sigma^x_m \sigma^x_{m+1}-1 \right)\right]  = |+ \rangle \langle +| + |- \rangle \langle -| 
\label{eq:ap5}
\end{equation}
where 
\begin{equation}
\sigma^x_m |\pm \rangle = \pm |\pm \rangle, \ \ m=1, \ldots , M
\label{eq:ap6}
\end{equation}
Given that the $| \pm \rangle$ are also eigenvectors of ${ \bf V}_2$, we are led to another version of (\ref{eq:ap4}):
\begin{eqnarray}
\lefteqn{Z_h} \nonumber \\ & = & ( 2 \sinh 2K_1)^{MN/2} e^{-MK_0} \left \{\langle +| { \bf V}^N | + \rangle + \langle - | { \bf V}^N |- \rangle \right\} \nonumber \\
\label{eq:ap7}
\end{eqnarray}
where
\begin{equation}
{ \bf V}= { \bf V}_2^{1/2} { \bf V}_1 { \bf V}_2^{1/2}
\label{eq:ap8}
\end{equation}
Finally, it is straightforward to verify that for periodic (or toroidal) boundary conditions
\begin{equation}
Z_t= (2 \sinh 2K_1)^{MN/2} \mathop{\rm Tr} { \bf V}^N
\label{eq:ap9}
\end{equation}

It is instructive to consider what can be deduced by duality. We can avoid having to consider the effect on eigenvectors by considering (4.8) and (4.14) and examine the pair of operator products \emph{before} taking the trace or the limit. A unitary transformation of the Fermi operators can be constructed \cite{DBA} which interchanges $\sum_1^M(- \sigma^z_j)$ and $\sigma_1^M\sigma_j^x \sigma_{j+1}^x$; there is a technical problem involving the boundary, which has been overcome elsewhere \cite{DBA}. This converts (4.8) into (4.15), but with $K_0$   replaced by $K_0^*$. The limit goes through and (4.17) is identical up to numerical factors, provided $K_1^*$ and $K_2$  are interchanged. Thus, if we have evaluated the right hand side of one or other of these two equations, the other follows directly. Duality in the usual form is discussed in the appendix. What we are using here is an \emph{operator } version of it as an automorphism of the algebra of Fermi operators which appears to have been known to Onsager, although this work was not published, as far as we know.

\subsection{The LSM approach}

In the LSM method Fermi operators $f_m$, $f_k^{\dagger}$ are introduced by means of the Jordan-Wigner transformation \cite{JW}
\begin{equation}
f_k = P_{j-1} \sigma_j^-, \ \ 2 \leq j \leq M \ \mbox{and} \ f_1= \sigma_1^-
\label{eq:a7}
\end{equation}
where $\sigma_j^-$ is the spin lowering operator, and $P_j$ is defined by
\begin{equation}
P_j = \prod_{l=1}^j (- \sigma_j^z), \ \ j=1, \ldots, M
\label{eq:a8}
\end{equation}
It is easy to check that the $f_j$ and $f_j^{\dagger}$ are, indeed, Fermi operators, and that
\begin{equation}
{ \bf V}_1 =  \exp \left[ -K_1^* \sum_{j=1}^M(2f_j^{\dagger}f_j-1)\right]
\label{eq:a9}
\end{equation}
The operator ${ \bf V}_2$ has a more subtle form:
\begin{eqnarray}
{ \bf V}_2
& = &\exp\left[ K_2 \Bigg\{ \sum_{j=1}^{M-1}(f_j^{\dagger} -f_j)(f_{j+1}^{\dagger} +f_{j+1}) \right. \nonumber \\
&& \left. - P_M(f^{\dagger}_M-f_M)(f_1^{\dagger}+f_1) \Bigg\} \right]
\label{eq:a10}
\end{eqnarray}
One can verify that
\begin{equation}
\left[ { \bf V}_{1,2},P_M\right] =0
\label{eq:a11}
\end{equation}
so that, for instance, ${ \bf V}^{\prime}$ and $P_M$ can be diagonalized simultaneously. The operator $P_M$ has eigenvalues $\pm 1$. Thus, we can write
\begin{equation}
{ \bf V}^{\prime} = { \bf P}_+ { \bf V}^{\prime}(+) + { \bf P}_- { \bf V}^{\prime}(-)
\label{eq:a12}
\end{equation}
where
\begin{equation}
{ \bf V}^{\prime}(\pm) = { \bf V}_1^{1/2}{ \bf V}_2(\pm) { \bf V}_1^{1/2}
\label{eq:a13}
\end{equation}
and
\begin{eqnarray}
{ \bf V}_2(\pm) & = & \exp \left[ K_2 \Bigg \{  \sum_{j=1}^{M-1} (f_j^{\dagger} -f_j)(f_{j+1}^{\dagger} + f_{j+1}) \right. \nonumber \\ && \left.  \mp (f_M^{\dagger} -f_M)(f_1^{\dagger}+f_1) \Bigg \} \right]
\label{eq:a14}
\end{eqnarray}
The operators ${ \bf P}_{\pm}$ are projectors onto the eigenstates of $P_M$. 

Since $P_M |0 \rangle = |0 \rangle$, it follows that
\begin{equation}
Z_f = (2 \sinh(2K_1))^{M(N-1)/2} 2^M e^{-MK_1^*}\langle 0 | ({ \bf V}^{\prime}(+))^N |0 \rangle
\label{eq:a15}
\end{equation}
It should be noted that if a line of reversed $K_2$ bonds were to be inserted on every tier of the cylindrical lattice between site $M$ and site 1, then the resultant partition function, which we denote $Z_f^a$, is given by
\begin{equation}
Z_f^a = (2 \sinh(2K_1))^{M(N-1)/2} 2^M e^{-MK_1^*}\langle 0|({ \bf V}^{\prime}(-))^N|0 \rangle
\label{eq:a16}
\end{equation}

The next step is to introduce running wave vectors as follows:
\begin{equation}
F^{\dagger}(k) = M^{-1/2} \sum_{j=1}^M e^{ikj}f_j^{\dagger}
\label{eq:a17}
\end{equation}
Both periodic (for the minus sign) and antiperiodic (for the plus sign) wave numbers are required to reduce (\ref{eq:a14}) to the form
\begin{eqnarray}
{ \bf V}_2(\pm)& =&  \exp \Bigg[K_2 \sum_{k \in \Omega_M( \pm)}e^{ik} (F^{\dagger}(k) - F(-k)) \nonumber \\ && (F^{\dagger}(-k) + F(k)) \Bigg]
\label{eq:a18}
\end{eqnarray}
where the sums over $k$ are to be taken in the range
\begin{eqnarray}
\Omega_M(\pm)=  \left\{ k: \exp [iMk] = \mp 1, \ - \pi < k \leq \pi \right\}
\label{eq:a19}
\end{eqnarray}
Equation (\ref{eq:a18}) can be analyzed further by grouping equal and opposite $k$ values together, giving
\begin{equation}
{ \bf V}_2(\pm) = \exp \left[2K_2 \sum_{0 \leq k < \pi} \left\{  \tau^z(k) \cos k +  \tau^y(k) \sin k \right\} \right]
\label{eq:a20}
\end{equation}
where
\begin{equation}
\tau^z(k) = (F^{\dagger}(k) F(k) + F^{\dagger}(-k) F(-k) -1)
\label{eq:a21}
\end{equation}
and
\begin{equation}
\tau^y(k) = -i \{F^{\dagger}(-k) F^{\dagger}(k) - F(k) F(-k)  \}
\label{eq:a22}
\end{equation}
Terms in (\ref{eq:a20}) for different values of the summand commute. In the paired subspaces, the $\tau^{\alpha}(k), \ \alpha=x,y,z$ are indeed spin $1/2$ operators. Thus, we can write ${ \bf V}_2(\pm)$ as a product
\begin{equation}
{ \bf V}_2(\pm) = \prod_{0 \leq k < \pi} \exp \left[2  K_2\left\{\tau^z(k) \cos k +  \tau^y(k) \sin k \right\} \right]
\label{eq:a23}
\end{equation}
Similarly, for the operator ${ \bf V}_1$
\begin{equation}
{ \bf V}_1(\pm) = \prod_{0 \leq k < \pi} \exp \left[ -2K_1^{*} \tau^z(k)\right]
\label{eq:a24}
\end{equation}
Thus, ${ \bf V}^{\prime}(\pm)$ will also factorize:
\begin{equation}
{ \bf V}^{\prime}(\pm) = \prod_{0 \leq k < \pi}{ \bf V}^{\prime}(k)
\label{eq:a25}
\end{equation}
where the $k$ values are determined according to (\ref{eq:a19}). The behavior at $k=0$ and $k=\pi$ must be specified carefully:
\begin{equation}
{\bf{V'}}\left( 0 \right) = \exp \left[ \left( {K_2  - K_1^ *  } \right)\left( {2F^\dag  \left( 0 \right)F\left( 0 \right) - 1} \right) \right]
\label{eq:a26}
\end{equation}
and
\begin{equation}
{\bf{V'}}\left( \pi  \right) = \exp \left[ - \left( {K_2  + K_1^ *  } \right)\left( {2F^\dag  \left( \pi  \right)F\left( \pi  \right) - 1} \right) \right]
\label{eq:a27}
\end{equation}
Then, the eigenstates of these operators are easily constructed. For the remaining ${ \bf V}^{\prime}(k)$, note first that
\begin{equation}
{ \bf V}^{\prime}(k) F^{\dagger}( \pm k) |0\rangle = F^{\dagger}( \pm k) |0 \rangle
\label{eq:a28}
\end{equation}
In the subspace with vectors $|0 \rangle$ and $F^{\dagger}(-k) F^{\dagger}(k) |0 \rangle$, the operators $\tau^{y,z}(k)$ introduced in (\ref{eq:a21}) and (\ref{eq:a22}), along with the operator
\begin{equation}
\tau^x(k) = F^{\dagger}(-k) F^{\dagger}(k) + F(k) F(-k)
\label{eq:a29}
\end{equation}
behave like Pauli spin operators, as was observed in \cite{SML}. This shows that for $k \neq 0, \pi$,
\begin{eqnarray}
\lefteqn{{\bf{V'}}\left( k \right)} \nonumber \\  &=&  \cosh (\gamma(k)) \nonumber  - \sinh( \gamma(k))  \nonumber \\ && \times\left( \cosh( \delta^{\prime}(k)) \tau^z(k) - \sin( \delta^{\prime}(k)) \tau^y(k)  \right)
\label{eq:a30}
\end{eqnarray}
where the function $\gamma(k)$ is determined by
\begin{eqnarray}
\cosh ( \gamma(k)) \ &=&  \cosh (2K_1^{*}) \cosh (2K_2) -  \nonumber  \\ && - \sinh(2K_1^*) \sinh (2K_2) \cos (k) \nonumber \\ 
\label{eq:a31}
\end{eqnarray}
where $\gamma(k)$ is also non-negative for real argument. The angle $\delta^{\prime}(k)$ is defined by
\begin{eqnarray}
\cosh( 2K_2) &=& \cosh (2K_1^*) \cosh (\gamma(k))  \nonumber \\ && - \sinh (2K_1^*) \sinh (\gamma(k)) \cos (\delta^{\prime}(k)) \nonumber \\ 
\label{eq:a32}
\end{eqnarray}
and 
\begin{equation}
\sin ( \delta ^{\prime}(k)) \sinh (\gamma(k)) = \sin k \sinh (2K_2)
\label{eq:a33}
\end{equation}
Notice that the term multiplying $\sinh( \gamma(k))$ in (\ref{eq:a30}) is itself a spin operator in the paired subspace. Raising (\ref{eq:a30}) to the $N^{\rm th}$ power is, then, straightforward:
\begin{eqnarray}
\lefteqn{({ \bf V}^{\prime}(k))^N} \nonumber \\ & = & \cosh(N \gamma(k)) \nonumber \\
&&- \sinh(N \gamma(k)) ( \cos( \delta^{\prime}(k)) \tau^z(k) - \sin ( \delta^{\prime}(k)) \tau^y(k)) \nonumber \\ 
\label{eq:a34}
\end{eqnarray}
Since $\tau^z(k) |0 \rangle = -|0 \rangle$ and for $M$ even, $\exp(i \pi M) =1$, evaluating the matrix element in (\ref{eq:a4}) is an easy matter: 
\begin{eqnarray}
\lefteqn{Z_f} \nonumber \\
& = & \exp [N(K_1^* + K_2)]  \nonumber \\ && \times \prod_{0<k< \pi} \left\{ \cosh(N \gamma(k)) + \sinh(N \gamma(k)) \cos (\delta^{\prime}(k))  \right\} \nonumber \\ 
\label{eq:a35}
\end{eqnarray}
Checking the behavior of $\delta^{\prime}( \pi)$, we see that this equation can be written as
\begin{equation}
Z_f = \prod_{k \in \Omega_M(+)} \{ \cos (N \gamma(k)) + \sinh(N \gamma(k)) \cos (\delta^{\prime}(k)) \}^{1/2}
\label{eq:a36}
\end{equation}
Thus
\begin{eqnarray}
\lefteqn{\ln Z_f} \nonumber \\& =& \frac{1}{2} \sum_{k \in \Omega_M(+)} \ln \left( \cosh( N \gamma(k)) + \sinh (N \gamma(k)) \cos( \delta^{\prime}(k)) \right) \nonumber \\ \label{eq:Zf}
\end{eqnarray}
This is in fact correct whether $M$ is odd or even.  The result for $Z_f^a$, the case in which antiperiodicity is enforced in each $M$-spin row, is a simple variation on (\ref{eq:Zf}). One simply replaces the requirement $k \in \Omega_M(+)$ in the sum by $k \in \Omega_M(-)$. 

\subsection{Periodic and antiperiodic boundary conditions}
\label{sec:perantiper}

The limit as $N \rightarrow \infty$ is easily taken, since $M < \infty$ implies that the sum is finite. Another way of examining this is to go back to (4.12). As $N \rightarrow \infty$, we expect only the contribution of the maximum eigenvector to survive, this provided its scalar product with $|0\rangle$  does not vanish, which turns out to be the case on detailed calculation. The summand in (1.45) becomes $N \gamma$,  and the sum may be implemented as a contour integral, following the methods of \cite{AS1,AS2}:
\begin{equation}
\lim_{N \rightarrow \infty}N^{-1} \ln Z_f = - \frac{M}{4 \pi} \oint_C \frac{d \omega}{1+e^{iM \omega}} \gamma( \omega) 
\label{eq:0.1}
\end{equation}
where $C$  is the rectangle in the complex $\omega$ plane with sides which  are segments of the lines $\mathop{\rm Re}(\omega) = \pm \pi, \mathop{\rm Im}(\omega) = \pm \epsilon$. The number $\epsilon$ is chosen small enough so that the only singularities of the integrand inside the contour are the zeros of the denominator. Equation (1.46) can then be written as:
\begin{eqnarray}
\lim_{N \rightarrow \infty} N^{-1} \ln Z_f  &=&  \frac{M}{4 \pi} \int_{-\pi}^{\pi} d \omega \  \gamma(\omega) \nonumber \\ 
&&- \frac{M}{2 \pi} \int_{- \pi + i \epsilon}^{\pi + i \epsilon} d \omega \frac{e^{iM \omega}}{1+ e^{i M \omega}} \gamma(\omega)\nonumber \\
\label{eq:0.2}
\end{eqnarray}
The first term on the right hand side is the bulk term. The Casimir contribution is contained in the second term on the right hand side of (\ref{eq:0.2}). The function $\gamma$ has branch cuts on the imaginary axis symmetrically positioned about the real axis. Using the $2 \pi$  periodicity of the integrand in (\ref{eq:0.2}), the contour in (\ref{eq:0.2}) will be deformed into one surrounding the branch cut in the upper half plane. It is convenient to make the transformation $\omega = i \hat{\gamma}(u)$ and interchange $K1$ and $K2$ in Eq. (4.44). Expanding the denominator, integrating by parts and re-summing gives the Casimir free energy from (\ref{eq:0.2}) as
\begin{equation}
F_{cas} = - \frac{k_BT}{\pi} \int_0^{\pi} du \ln \left[ 1 + \exp \left( - M \hat{\gamma(u)}\right) \right]
\label{eq:0.4}
\end{equation}
The resulting Casimir force is denoted $f_{cas}(cyl.)$ and is just:
\begin{equation}
f_{cas}(cyl.) = - \frac{k_BT}{4 \pi} \int_{-\pi}^{\pi} du \ \hat{\gamma}(u) \left[ 1 - \tanh \left(  M \hat{\gamma}(u)/2\right)\right]
\label{eq:0.5}
\end{equation}
Note that this expression is invariant under interchange of $K_1^*$  and $K_2$; that is, there is a dual symmetry; further, this force is {\em attractive}. The scaling form (with $K_1=K_2$ ) is accessed by noting that $\hat{\gamma}(u) \sim [\hat{\gamma}(0)^2+u^2]^{1/2}$  for small $u$  and $\hat{\gamma}(0)$. We then introduce scaling variables $y=Mu/2$   and $x=M \hat{\gamma}(0)/2$   and take the scaling limit, giving:
\begin{eqnarray}
M^2 \bar{f}_{cas}(cyl.) & \rightarrow &- \frac{k_BT}{\pi} \int_{-\infty}^{\infty} dy (x^2+y^2)^{1/2} \nonumber\\
&& \left[1- \tanh \left( (x^2+y^2)^{1/2}\right) \right]
\label{eq:0.6}
\end{eqnarray}
This implies the same scaling function as in (3.15).
The integral in (\ref{eq:0.6}) has the value $\pi^2/12$ at  $x=0$.

The evaluation of $Z_f^a$ is also quite straightforward with this procedure; the analogue of (\ref{eq:0.2}) is: 
\begin{eqnarray}
\lim_{N \rightarrow \infty} N^{-1}Z_f^a &=& \frac{M}{4 \pi} \int_{-\pi}^{\pi} d \omega \ \gamma(\omega) \nonumber\\
&& + \frac{M}{2 \pi} \int_{- \pi + i \epsilon}^{\pi + i \epsilon} d \omega \frac{e^{iM \omega}}{1- e^{i M \omega}} \gamma(\omega) \nonumber\\
\label{eq:0.7}
\end{eqnarray}
with an additional term $\mathop{\rm sgn} (T_c-T) \gamma(0)$,   which will not appear in the Casimir term, since it is independent of  $M$. Note that the only other difference is the sign in the denominator of the second integrand; this comes about because the sum leading to this term is over \emph{periodic} wave numbers. The Casimir free energy in this case is denoted by $F^a_{cas}(cyl.)$; it is given by:
\begin{equation}
F_{cas}^a(cyl.) = - \frac{k_BT}{\pi} \int_0^{\pi} du \ \ln\left[1-\exp \left( -M \hat{\gamma}(u)\right) \right]
\label{eq:0.8}
\end{equation} 
Note that this expression is also invariant under interchange of $K_1^*$  and  $K_2$.
The analogue of (\ref{eq:0.6}) for the Casimir force with the line of reversed bonds is:
\begin{eqnarray}
M^2 \bar{f}^a_{cas}(cyl.) &\rightarrow& \frac{k_BT}{\pi} \int_{\infty}^{\infty}dy \ (x^2+y^2)^{1/2} \nonumber\\
&&\left[\coth \left( (x^2+y^2)^{1/2}\right) - 1 \right]
\label{eq:0.9}
\end{eqnarray}
The integral for $x=0$  has the value  $\pi^2/6$.

\subsection{Free and fixed boundary conditions}
\label{sec:freefixed}

To calculate the Casimir force for free and fixed boundary conditions we consider the other limits, i.e. $M \rightarrow \infty$ with $N$ finite. This yields
\begin{eqnarray}
\lim_{M \rightarrow \infty} \frac{1}{M}\ln Z_f  &=& \frac{1}{4 \pi} \int_0^{2 \pi} \ln \bigg[\cosh(N \gamma(\omega) ) \nonumber \\ &&+ \sinh(N \gamma(\omega)) \cos \delta^{\prime}(\omega)  \bigg] \ d \omega \nonumber \\
\label{eq:ff1}
\end{eqnarray}
Extracting the bulk term we are left with an incremental part, denoted $F_{f}^{cas}$. This remainder is given by
\begin{eqnarray}
F_f^{cas} &=& - \frac{k_BT}{4 \pi} \int_{-\pi}^{\pi} \ln \bigg[(1+ \cos (\delta^{\prime}(\omega)))  \nonumber \\ && +(1- \cos (\delta^{\prime}(\omega))) \exp[-2N \gamma(\omega)] \bigg] \ d \omega \nonumber \\
\label{eq:ff2}
\end{eqnarray}
Taking the derivative of the above result with respect to $N$,  

\begin{eqnarray}
F_f^{cas} &=& - \frac{k_BT}{2 \pi} \int_{-\pi}^{\pi}  \gamma(\omega) \ d \omega \nonumber \\ 
&& \frac{1}{ 1+[(1+ \cos (\delta^{\prime}(\omega))) /(1- \cos (\delta^{\prime}(\omega)))] \exp[2N \gamma(\omega)]}, \nonumber \\
\label{eq:ff22}
\end{eqnarray}

\noindent and introducing scaled variables as in Section  \ref{page:sstart} with $M$ replaced by $N$, we obtain the following expression for the Casimir force
\begin{eqnarray}
\lefteqn{f_{cas}^{\rm free} } \nonumber \\ &= & -\frac{2}{\pi N^2} \int_{-\infty}^{\infty} d \Omega \sqrt{t^2+ \Omega^2} \nonumber \\
&&\left[\frac{(1-\frac{t}{\sqrt{t^2+ \Omega^2}})e^{-4\sqrt{t^2+\Omega^2}}}{(1+\frac{t}{\sqrt{t^2+ \Omega^2}})+(1-\frac{t}{\sqrt{t^2+ \Omega^2}})e^{-4\sqrt{t^2+\Omega^2}}} \right] \nonumber \\
\label{eq:ff3}
\end{eqnarray}
which agrees with the expression (\ref{eq:2bc15}) derived by other methods. 

An analogous argument allows us to evaluate the Casimir free energy for the strip with plus spins at both ends. The right hand side of (\ref{eq:ap5}) can be evaluated in a convenient form for calculations by relating it to the limit of { \bf V} when $K_1^* \rightarrow 0$. Following this with the limit $K_0 \rightarrow \infty$ then yields
\begin{eqnarray}
\lefteqn{\lim_{K_0\rightarrow \infty} \exp \left[ K_0 \sum_{m=1}^M \left(\sigma_m^x \sigma_{m+1}^x -1 \right)\right]} \nonumber \\ & =&  | \Phi_+^0 \rangle \langle \Phi_+^0| + | \Phi_-^0 \rangle \langle \Phi_-^0 |
\label{eq:ff4}
\end{eqnarray}
The states on the right hand side of (4.72) are defined by $\lim_{K_1^* \rightarrow 0}|\Phi_{\pm} \rangle = | \Phi_{\pm}^0 \rangle$.
Thus they are eigenvalues  of ${\bf V}_2(\pm)$ written in a form which is useful for the calculation.  Then a straightforward calculation gives
\begin{eqnarray}
\lefteqn{\langle \Phi_{\pm}^0 |  ({ \bf V}(\pm))^N | \Phi_{\pm}^0 \rangle } \nonumber \\ & = & \prod_{k \in \Omega_M(\pm)} \left[ \cosh(N \gamma(k)) + \sinh(N \gamma(k)) \cos \delta^*(k)\right]^{1/2} \nonumber \\
\label{eq:ff5}
\end{eqnarray}
where the angle $\delta^*$ is defined by (\ref{eq:a32}) and (\ref{eq:a33}) \emph{with} $K_1^*$ \emph{and} $K_2$ \emph{interchanged}, which is just what we could have anticipated from duality, as explained above. Evaluating the analogue of (\ref{eq:0.4}) we find
\begin{eqnarray}
\lefteqn{f^{ ++}_{cas}} \nonumber \\ & = & - \frac{2}{\pi N^2} \int_{-\infty}^{\infty} d \Omega \sqrt{t^2+ \Omega^2} \nonumber \\ &&\left[\frac{(1+\frac{t}{\sqrt{t^2+ \Omega^2}})e^{-4\sqrt{t^2+\Omega^2}}}{(1-\frac{t}{\sqrt{t^2+ \Omega^2}})+(1+\frac{t}{\sqrt{t^2+ \Omega^2}})e^{-4\sqrt{t^2+\Omega^2}}} \right] \nonumber \\
\label{eq:ff6}
\end{eqnarray}
Thus 
\begin{equation}
f^{++}_{cas}(t) = f^{\rm free}_{cas}(-t)
\label{eq:ff7}
\end{equation}
which is a manifestation of duality. 

\section{conclusion}
\label{conclusion}

  In this paper, we exploit the exact solutions for the partition function of the Ising model in one and two dimensions to examine the impact of boundary conditions and dimensionality on the Casimir forces. Our results are effectively summarized in Figs. \ref{fig:apGplot} and \ref{fig:compare3}. As both figures unambiguously illustrate, boundary conditions alone do not suffice to predict the relative strengths of critical Casimir forces. This is perhaps most strikingly evident in the case of the forces associated with periodic and free boundary conditions. While free boundary conditions give rise to a vanishing critical Casimir force in one dimension, in the two dimensional case the same boundary conditions result in a significant attractive force. By contrast, periodic boundary conditions yield a non-zero critical Casimir force in one dimension--that is, a force that is infinitely larger than the one associated with free boundary conditions. On the other hand, in two dimensions, the Casimir force generated by the same boundary conditions  is smaller in amplitude than the Casimir force resulting from free boundary conditions.

In the case of anti-periodic boundary conditions, the critical Casimir force is repulsive in both one and two dimensions. An additional contrast between periodic and antiperiodic  boundary conditions is evident in the limit $T=0$ in one dimension; at that point, the Casimir force is zero for the case of periodic boundary but is finite in the anti-periodic case.
 
 The methods presented in this paper can also be extended to calculate Casimir forces with mismatched boundary conditions.

\begin{acknowledgments}
We are pleased to acknowledge useful and important discussions with Professor Mehran Kardar. Financial support from NSF grant numbers DMR-06-45668 and DMR-07-04274 is greatly acknowledged.
\end{acknowledgments}

\pagebreak
\appendix

\section{Derivation of exact partition functions for the Ising Model with antiperiodic-antiperiodic boundary conditions}
\label{app:wuhu}
Here, we make use of results in Ref. \cite{wuandhu} for the reduced partition function of the two dimensional square Ising model with antiperiodic boundary conditions in both directions (see Eq. (79) in Ref. \cite{wuandhu} )
\begin{equation}
Q^{aa}=  \frac{1}{2} \left[ -\Omega_{\frac{1}{2}, \frac{1}{2}}+  \Omega_{\frac{1}{2}, 0}+  \Omega_{0, \frac{1}{2}} + \mathop{ \rm sgn} \left(\frac{\theta - \theta_c}{\theta_c} \right)  \Omega_{0,0}\right]
\label{eq:apap1}
\end{equation}
with  $\theta = k_BT/J$
and 
\begin{widetext}
\begin{eqnarray}
\lefteqn{\Omega_{\mu, \nu}} \nonumber \\ & = & \prod_{p=0}^{N-1} \prod_{q=0}^{M-1} \Bigg[A_0-A_1 \cos \frac{2 \pi(p+\mu)}{N} -A_2 \cos \frac{2 \pi(q+\nu)}{M}    -A_3 \cos \left( \frac{2 \pi(p+\mu)}{N} -\frac{2 \pi(q+\nu)}{M} \right)\Bigg]^{1/2}
\label{eq:Ommunu}
\end{eqnarray}
\end{widetext}
Given 
\begin{equation}
t = \tanh \beta J
\label{eq:teq}
\end{equation}
for a square lattice with uniform interaction strength $J$, we find 
\begin{eqnarray}
A_0 & = & (1+t^2)^2 \label{eq:A01} \\
A_1 & = & 2t(1-t^2) \label{eq:A11} \\
A_2 & = & 2t(1-t^2), \label{eq:A21} \\
A_3 & = & 0 \label{eq:A31}
\end{eqnarray}
and thus,
\begin{widetext}
\begin{eqnarray}
\lefteqn{\Omega_{\mu \nu}} \nonumber \\ & = & \prod_{p=0}^{N-1} \prod_{q=0}^{M-1} \left[(1+t^2)^2 -2t(1-t^2) \cos \frac{2 \pi (p+\mu)}{N} -2t(1-t^2) \cos \frac{2 \pi (q+\nu)}{M} \right]^{1/2}\nonumber \\
\label{eq:Omegamunu}
\end{eqnarray}
\end{widetext}

The quantity $\Omega_{0,0}$ in Eq. (\ref{eq:Ommunu}) needs special attention.  Consider the $p=q=0$ term in the product for $\Omega_{0,0}$,
\begin{eqnarray}
[(1+t^2)^2 - 4t(1-t^2)]^{1/2}& =& [(1-2t-t^2)^2]^{1/2} \nonumber \\
& = & |1-2t-t^2|
\label{eq:Om001}
\end{eqnarray}
Making use of (\ref{eq:teq}), we have for the last line of (\ref{eq:Om001}) 
\begin{eqnarray}
\lefteqn{|1- 2 \tanh \beta J -\tan ^2 \beta J|} \nonumber \\ & = & \frac{1}{\cosh ^2 \beta J} | \cosh ^2 \beta J - \sinh ^2 \beta J - 2 \cosh \beta J \sinh \beta J | \nonumber \\ &= &\frac{1}{\cosh^2 \beta J}|1-\sinh 2 \beta J   |
\label{eq:Om002}
\end{eqnarray}
According to (\ref{eq:critK}),the critical temperature is at 
\begin{equation}
\cosh \beta J \coth \beta J =2,
\label{eq:critJ1}
\end{equation}
which can be rewritten as 
\begin{equation}
\sinh 2 \beta J = 1
\label{eq:critJ2}
\end{equation}
Thus, the $p=q=0$ term in the product for $\Omega_{00}$ is proportional to $| \theta - \theta_c|$. The term going as the sign of that expression guarantees that the product goes as $\theta - \theta_c$. It is a straightforward exercise to verify that all other factors in the product are free of singularities in the vicinity of the critical point. 

\subsection{Sum over the variable $q$}
Now we focus on the sum over the variable $q$ in Eq. (\ref{eq:Omegamunu}). There are two cases of interest: $\nu =0$ and $\nu=1/2$. We will analyze the case $\nu=0$ making use of the method outlined in Section \ref{sec:2danti}.  The generic sum of interest is
\begin{equation}
\sum_{q=0}^{M-1}\frac{1}{a-b \cos \frac{2 \pi q}{M}},
\label{eq:gensum1}
\end{equation}
which is equivalent to the integral
\begin{widetext}
\begin{eqnarray}
\lefteqn{\frac{1}{2 \pi i}\int \frac{1}{a-b(z+1/z)/2} \ \frac{Mz^{M-1}}{z^M-1}dz} \nonumber \\ &=& -\frac{1}{2 \pi i} \frac{2}{b}\int \frac{1}{z^2 - 2az/b +1}\frac{Mz^{N}}{z^M-1} dz \nonumber \\
& = & -\frac{1}{2 \pi i}\frac{2}{b} \int \frac{1}{(z-a/b-\sqrt{(a/b)^2-1})(z-a/b+\sqrt{(a/b)^2-1)}} \nonumber \\ && \times \frac{z^M}{z^M-1} dz
\label{eq:gensum2}
\end{eqnarray}
\end{widetext}
The integral is around the contours shown in Fig. \ref{fig:contours}. This integral is evaluated by closing it around the two poles in the first term in the intregrand on the third line of (\ref{eq:gensum2}). We find after evaluating residues that the result of this integration is
\begin{eqnarray}
\lefteqn{\frac{M}{\sqrt{a^2-b^2}}\left(\frac{(a/b+\sqrt{(a/b)^2-1})^M}{(a/b+\sqrt{(a/b)^2-1})^M-1} \right. } \nonumber \\ && \left. - \frac{(a/b-\sqrt{(a/b)^2-1})^M }{(a/b-\sqrt{(a/b)^2-1})^M-1} \right)
\label{eq:gensum3}
\end{eqnarray}
Setting $a/b+\sqrt{(a/b)^2-1} = \Gamma$ and using the fact that $a/b-\sqrt{(a/b)^2-1}=(a/b+\sqrt{(a/b)^2-1})^{-1}$, we can write (\ref{eq:gensum3}) as follows
\begin{eqnarray}
\lefteqn{\frac{M}{\sqrt{a^2-b^2}}\frac{\Gamma^M+1}{\Gamma^M-1}} \nonumber \\  & = & \frac{M}{\sqrt{a^2-b^2}}\frac{\Gamma^{M/2}+\Gamma^{-M/2}}{\Gamma^{M/2}- \Gamma^{-M/2}} \nonumber \\
& = & \frac{M}{\sqrt{a^2-b^2}}\coth\left(\frac{M}{2}\ln \Gamma\right)
\label{eq:gensum6}
\end{eqnarray}
The sum of interest is in fact,
\begin{equation}
\sum_{q=0}^{M-1} \ln \left(a-b \cos \frac{2 \pi q}{M} \right)
\label{eq:gensum7}
\end{equation}
which can be obtained by integrating the sum in (\ref{eq:gensum1}) with respect to $a$. Making use of the fact that $\frac{d}{da} \ln \left( a/b +\sqrt{(a/b)^2-1}\right) = 1/\sqrt{a^2-b^2}$, we find the antiderivative of the last line of (\ref{eq:gensum6}) with respect to $a$
\begin{equation}
2  \ln \left[ \sinh \left(\frac{M}{2} \ln \Gamma \right) \right]
\label{eq:gensum9}
\end{equation}
Finally, if we define $\gamma_p = \ln \Gamma_p$ we have a result that begins to look like a contribution to the expressions utilized by Ferdinand and Fisher in the case of periodic boundary conditions. To complete the connection, we note that in this case
\begin{eqnarray}
a & = & (1+t^2)^2-2t(1-t^2) \cos \frac{2 \pi(p+\mu)}{N} \label{eq:adef1} \\
b & = & 2t(1-t^2) \label{eq:bdef1}
\end{eqnarray}
Then,
\begin{eqnarray}
\lefteqn{\Gamma_p} \nonumber \\ & = & \frac{(1+t^2)^2}{t(1-t^2)} - \cos \frac{2 \pi(p+\mu)}{N}  \nonumber \\ && + \sqrt{\left( \frac{(1+t^2)^2}{t(1-t^2)} - \cos \frac{2 \pi(p+\mu)}{N} \right)^2 -1} \nonumber\\
& = & \cosh 2 \beta J \coth 2 \beta J - \cos \frac{2 \pi(p+\mu)}{N} \nonumber \\ &&+\sqrt{\left(  \cosh 2 \beta J \coth 2 \beta J - \cos \frac{2 \pi(p+\mu)}{N}\right)^2-1} \nonumber \\
\label{eq:Gammaresult}
\end{eqnarray}
Eq. (\ref{eq:gensum9}) corresponds to the logarithm of the product over $q$ of $\Omega_{\mu 0}$.  Reconstituting the product over $p$, we have
\begin{equation}
\Omega_{\mu 0} = \prod_{p=0}^{N-1}  \sinh \left(\frac{M}{2} \gamma_p \right)
\label{eq:Ommu0}
\end{equation}
Let's focus on one term in the product for $\Omega_{00}$: the $p=0$ term. It is
\begin{eqnarray}
\lefteqn{\sinh \Bigg(\frac{M}{2} \ln \Bigg(\cosh 2 \beta J \coth 2 \beta J - 1+  } \nonumber \\ &&   \sqrt{\left(\cosh 2 \beta J \coth 2 \beta J - 1\right)^2-1} \Bigg)  \Bigg)
\label{eq:term1a}
\end{eqnarray}
The function $\cosh 2 \beta J \coth 2 \beta J $ has a minimum at $\beta J = \cosh ^{-1}\left(\sqrt{\frac{1}{2} \left(1+\sqrt{2}\right)}\right)$. We write
\begin{equation}
\beta J  = \cosh ^{-1}\left(\sqrt{\frac{1}{2} \left(1+\sqrt{2}\right)}\right) + \tau
\label{eq:expandbetaJ}
\end{equation}
Then, the argument of the hyperbolic sine in (\ref{eq:term1a}) becomes
\begin{equation}
\frac{M}{2} \ln \left[1+4 \tau+8 \tau^2+O\left(\tau^3\right) \right] = 2 M \tau
\label{eq:sinharg}
\end{equation}
which means that as the temperature passes through the critical temperature, the hyperbolic sine function changes sign. This is the source of the ``sgn'' term in (\ref{eq:part3}) and (\ref{eq:apap1}). 

The calculations utilized to obtain $\Omega_{\mu 0}$ can be adapted to the evaluation of the sum over $q$ in $\Omega_{\mu \frac{1}{2}}$. After a set of tedious but straightforward calculations, we obtain for that result
\begin{equation}
\Omega_{\mu \frac{1}{2}} = \prod_{p=0}^{N-1} \cosh \left( \frac{M}{2} \gamma_p \right)
\label{eq:Omegamu1/2}
\end{equation}

\section{Contributions of the terms in (\ref{eq:part3}) to the partition function in the case of antiperiodic boundary conditions} \label{sec:details}

\begin{figure}[htbp]
\begin{center}
\includegraphics[width=3in]{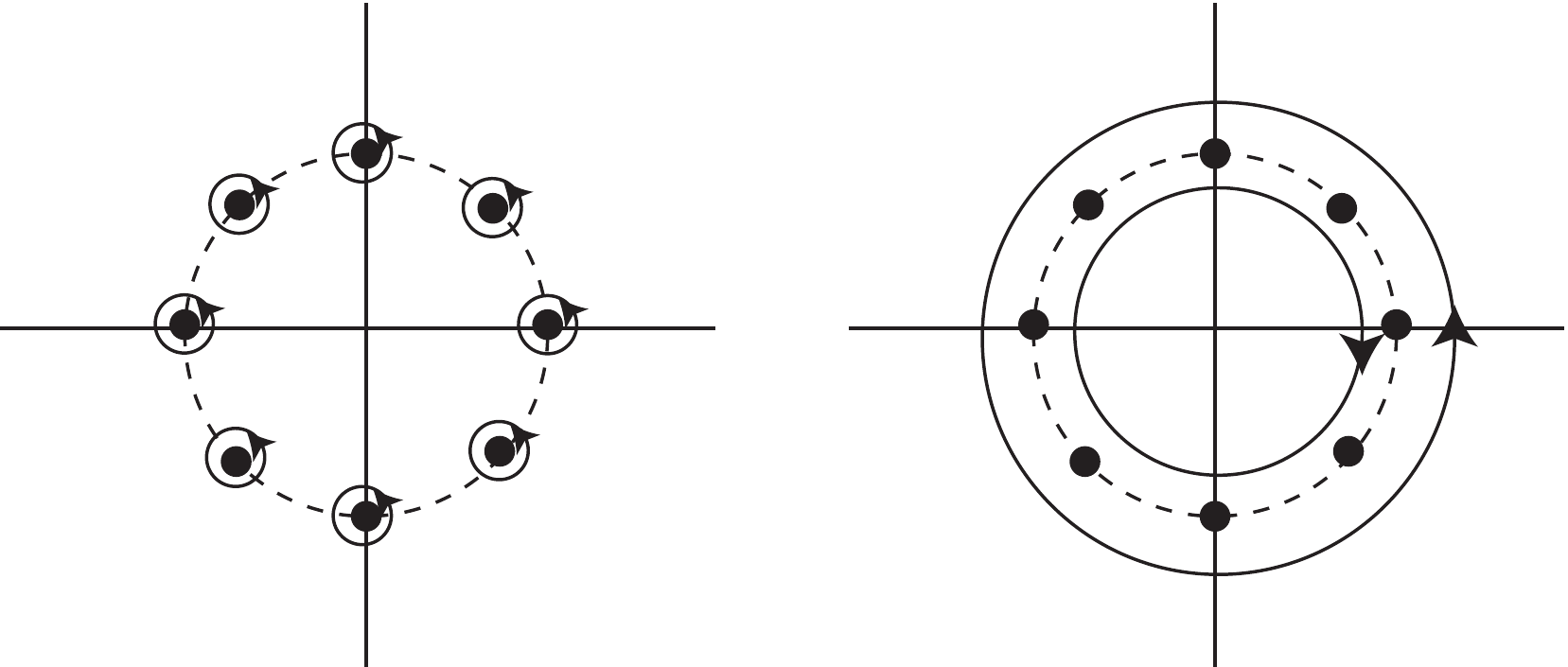}
\caption{Contours for the integral on the right hand side of (\ref{eq:cancel3a}). }
\label{fig:contours}
\end{center}
\end{figure}

We first focus on the sums involving hyperbolic cosines. We have
\begin{equation}
\sum_{r=1}^{N-1} \ln \cosh \frac{1}{2} M \gamma_{2r+1} = \frac{1}{2 \pi i} \int \ln \cosh \frac{M}{2} \gamma(z) \frac{Nz^{N-1}}{z^N +1} dz
\label{eq:cancel2a}
\end{equation}
and
\begin{equation}
\sum_{r=1}^{N-1} \ln \cosh \frac{1}{2} M\gamma_{2r} = \frac{1}{2 \pi i} \int \ln \cosh \frac{M}{2} \gamma(z) \frac{Nz^{N-1}}{z^N-1} dz
\label{eq:cancel3a}
\end{equation}
The contours of integration in (\ref{eq:cancel3a}) shown in Fig. \ref{fig:contours}; the contours of integration in (\ref{eq:cancel2a}) are identical except for a shift along the unit circle.  The contours enclose the roots of the function $1/(z^N-1)$, as depicted on the left hand side of that figure. They are then deformed to lie just inside and outside the unit circle---see the right hand side of Fig. \ref{fig:contours}.  We have already established in Section \ref{sec:periodic2d} that the other singularities in the integrands in (\ref{eq:cancel2a}) and (\ref{eq:cancel3a}) lie along the real axis. They are branch cuts which are close to the unit circle in the vicinity of the critical temperature. 

We now concentrate on the asymptotic expression in the vicinity of the critical point and perform expansions like those in Section \ref{sec:periodic2d} and in Eq. (\ref{eq:sinharg}) and find the following expression
\begin{equation}
\sum_{\phi_p} \ln \cosh \frac{M}{2}\sqrt{4\tau^2 + \phi_p^2},
\label{eq:cancel1}
\end{equation}
with $\phi_p = 2 \pi(p+1/2)/N$ in case of Eq. (\ref{eq:cancel2a}) and with $\phi_p = 2 \pi p/N$ for Eq. (\ref{eq:cancel3a}).

\begin{figure}[htbp]
\begin{center}
\includegraphics[width=3in]{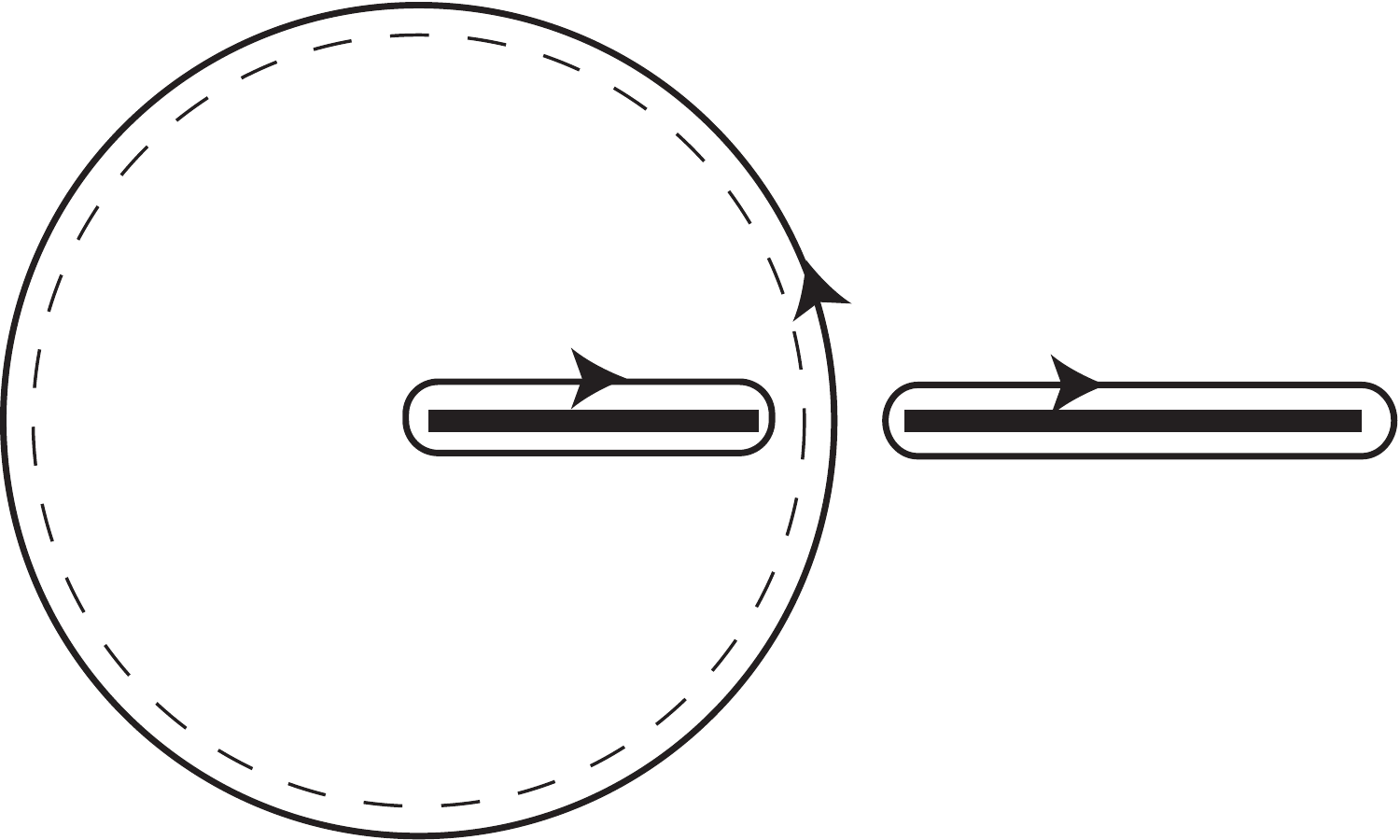}
\caption{The three contours over which integrations are performed. The dashed circle is the unit circle, and the circular contour actually lies right on the unit circle. }
\label{fig:cancellation1}
\end{center}
\end{figure}

This holds when the complex variable $z$ is in the immediate vicinity of $z=1$, where the unit circle passes through the real axis. In fact,  $\phi =\pm i(z-1)$. The branch points on the real axis are associated with the values of $z$ for which the argument of the natural logarithm goes through zero. This occurs when 
\begin{equation}
\cosh\frac{M}{2} \sqrt{4\tau^2+ \phi^2} =0
\label{eq:bp1}
\end{equation}
We cause the hyperbolic cosine to go through zero by making its argument imaginary so that it becomes the trigonometric cosine. To do this, we have $\phi \rightarrow \pm i \Phi$. When $\Phi>4\tau$, 
\begin{equation}
\cosh \frac{M}{2}\sqrt{4\tau^2 + \phi^2} \rightarrow \cos \frac{M}{2} \sqrt{\Phi^2- 4\tau^2}
\label{eq:bp2}
\end{equation}
The right hand side of (\ref{eq:bp2}) goes to zero when
\begin{equation}
\frac{M}{2} \sqrt{\Phi^2- 4\tau^2} = \frac{\pi}{2}
\label{eq:cancel1b}
\end{equation}
Solving for $\Phi$, we find
\begin{equation}
\Phi= \sqrt{\left(\frac{\pi}{M} \right)^2 + 4\tau^2}
\label{eq:cancel2b}
\end{equation}
Of course, the cosine goes to zero at an infinite set of points on the real axis, corresponding to $(M/2)\sqrt{\Phi^2-4\tau^2} =(2n+1) \pi/2$. That is to say, there is an infinite number of branch points on the real axis. However, we are interested in the one closest to the unit circle, corresponding to the conditions in (\ref{eq:cancel1b}) and (\ref{eq:cancel2b}). The two branch points of interest are just inside and just outsize the unit circle. The contribution of the integral around the branch cut ending at those branch points will be dominated by a multiplicative exponential factor. The portion of the distorted contour that lies just inside the unit circle can be pushed towards the origin so that it wraps around the branch cut. The general strategy for deforming the contours is illustrated in Fig. \ref{fig:cancellation1}.
Note that the branch cuts are along the real axis. To assess the contribution of the contour lying just outside the unit circle, we rewrite the last portions of the integrands as follows
\begin{eqnarray}
\frac{Nz^{N-1}}{z^N+1} & = & \frac{N}{z} -\frac{N}{z}\frac{1}{z^N+1} \label{eq:cancel4a} \\
\frac{Nz^{N-1}}{z^N-1}& = & \frac{N}{z} +\frac{N}{z}\frac{1}{z^N-1} \label{eq:cancel5a}
\end{eqnarray}
Therefore, the right hand side of (\ref{eq:cancel2a}) becomes
\begin{eqnarray}
\lefteqn{\frac{1}{2 \pi i} \int \ln \cosh \frac{M}{2} \gamma(z) \frac{N}{z} dz} \\ && - \frac{1}{2 \pi i} \int \ln \cosh \frac{M}{2} \gamma(z) \frac{N}{z}\frac{1}{z^N+1}  dz 
\label{eq:cancel6a}
\end{eqnarray}
and that of (\ref{eq:cancel2b}) becomes
\begin{eqnarray}
\lefteqn{\frac{1}{2 \pi i} \int \ln \cosh \frac{M}{2} \gamma(z) \frac{N}{z} dz } \nonumber \\ &&- \frac{1}{2 \pi i} \int \ln \cosh \frac{M}{2} \gamma(z) \frac{N}{z}\frac{1}{z^N-1}  dz 
\label{eq:cancel6f}
\end{eqnarray}
The first terms in (\ref{eq:cancel6a}) and (\ref{eq:cancel6f}) are integrated around the unit circle and become the dominant contribution to the sum we are performing. The second terms are deformed to wrap around the branch cut that is outside the unit circle. 

For the branch cut just outside the unit circle, $z=1+ \Phi$, and the factor of interest goes as 
\begin{eqnarray}
z^{-N}  & = & (1+ \Phi)^{-N} \nonumber \\
& \rightarrow & e^{-N \sqrt{(\pi/M)^2 + \Delta^2}}
\label{eq:cancel4b}
\end{eqnarray}
In the above, we have neglected the difference between factors going as $(1 \pm \Phi)$ and 1. This is because $\Phi$ is going to be small ($O(1/M)$). 

Looking at the contour around the branch cut outside the unit circle, we see that the magnitude and the sign of the contribution is fundamentally the same for both of the $\ln \cosh $ sums. On the other hand, because of the sign differences in the second terms on the right hand sides of (\ref{eq:cancel4a}) and (\ref{eq:cancel5a}), we see that the two sums will differ by exponentially small amounts, the size of the difference being dominated by $e^{-N \sqrt{(\pi/M)^2 + 4\tau^2}}$. 
In the case of the branch cut just inside the unit circle, $z=1- \Phi$, and we have to consider the factor
\begin{eqnarray}
z^N & = & (1- \Phi)^N \nonumber \\
& = & e^{N \ln (1- \Phi)} \nonumber \\
& \rightarrow & e^{-N \Phi} \nonumber \\
& = & e^{-N \sqrt{(\pi/M)^2 + \Delta^2}}
\label{eq:cancel3b}
\end{eqnarray}
Exponentiating the sums and expanding the exponentials with respect to the very small (in fact infinitesimally small in the thermodynamic limit $N \rightarrow \infty$) terms, we find that there is a remainder after the subtraction that goes as 
\begin{eqnarray}
\lefteqn{\exp \Bigg[- N\sqrt{4\tau ^2 + (\pi/M)^2}} \nonumber \\ &&  +\frac{N}{2 \pi} \int  \left(\ln \cosh \frac{M}{2} \sqrt{4\tau^2 + \phi^2}   \right) \ d \phi \Bigg]
\label{eq:cancel5b}
\end{eqnarray}
The second term in the exponential comes from the integral around the unit circle. There are also multiplicative terms, and they may depend non-trivially on key variables. However, they have only marginal effect on the overall amplitude of the expression in (\ref{eq:cancel5b}). 

In the case of the hyperbolic sine sums, the analysis is basically the same, up to the point at which we look for the branch point. Here, we look for the values of $z$ at which the hyperbolic sine goes through zero. This happens when
\begin{equation}
\sin \frac{M}{2} \sqrt{\Phi^2 - 4\tau^2} =0
\label{eq:cancel6b}
\end{equation}
The value of $\Phi$ for which this occurs that is closest to the intersection between the unit circle and the real axis ($z=1$) is
\begin{equation}
\Phi = \pm 2\tau
\label{eq:cancel7b}
\end{equation}
The same set of steps taken above leads us to a remainder if the exponentiated hyperbolic sine sums are subtracted from each other that goes as 
\begin{equation}
\exp \left[ -N|4\tau |  + \frac{N}{2 \pi} \int  \left(\ln \sinh \frac{M}{2} \sqrt{4\tau^2 + \phi^2}   \right) \ d \phi \right]
\label{eq:cancel8b}
\end{equation}

\begin{figure}[htbp]
\begin{center}
\includegraphics[width=3in]{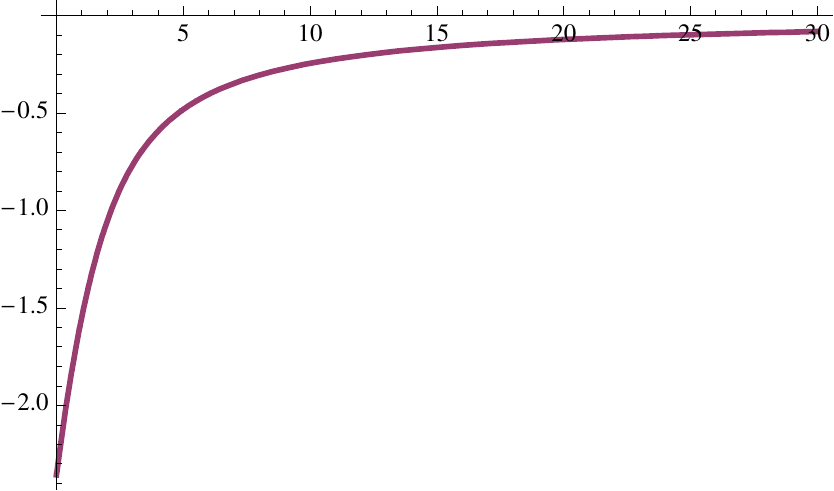}
\caption{The difference between the exponents in (\ref{eq:cancel7b}) and (\ref{eq:cancel8b}), plotted versus the scaled reduced temperature, $t$. }
\label{fig:cancelfig}
\end{center}
\end{figure}

The term in (\ref{eq:cancel8b}) dominates the term in (\ref{eq:cancel7b}). To see this, let's focus on the partition function in the region of interest, writing 
\begin{eqnarray}
\tau & = & \frac{t }{M} \label{eq:newDelta} \\
\phi & \rightarrow & 2 \frac{\Omega}{M} \label{eq:newphi}
\end{eqnarray}
Then, if we take the difference between the exponents in (\ref{eq:cancel7b}) and (\ref{eq:cancel8b}), we have
\begin{eqnarray}
\lefteqn{\frac{N}{M} \Bigg[ \frac{1}{\pi} \int_{-\infty}^{\infty} \ln \coth \sqrt{t^2 + \Omega^2} \ d \Omega} \nonumber \\ &&- \sqrt{\pi^2 + 4t^2  } + 2  |t| \Bigg]
\label{eq:cancel9b}
\end{eqnarray}
The plot of this difference is shown in Fig. \ref{fig:cancelfig}.
Note that it is always negative.

\section{Duality and boundary conditions} \label{app:duality}

Establishment of the duality between the two dimensional Ising lattice with free boundary conditions and the lattice with fixed boundary conditions can be achieved in two ways: by comparison of high and low temperature expansions for the partition function or with the use of automorphisms in the Kauffman operators. We will review both approaches in this Appendix

\subsection{Duality via high and low temperature expansions}

\begin{figure}[htbp]
\begin{center}
\includegraphics[width=2in]{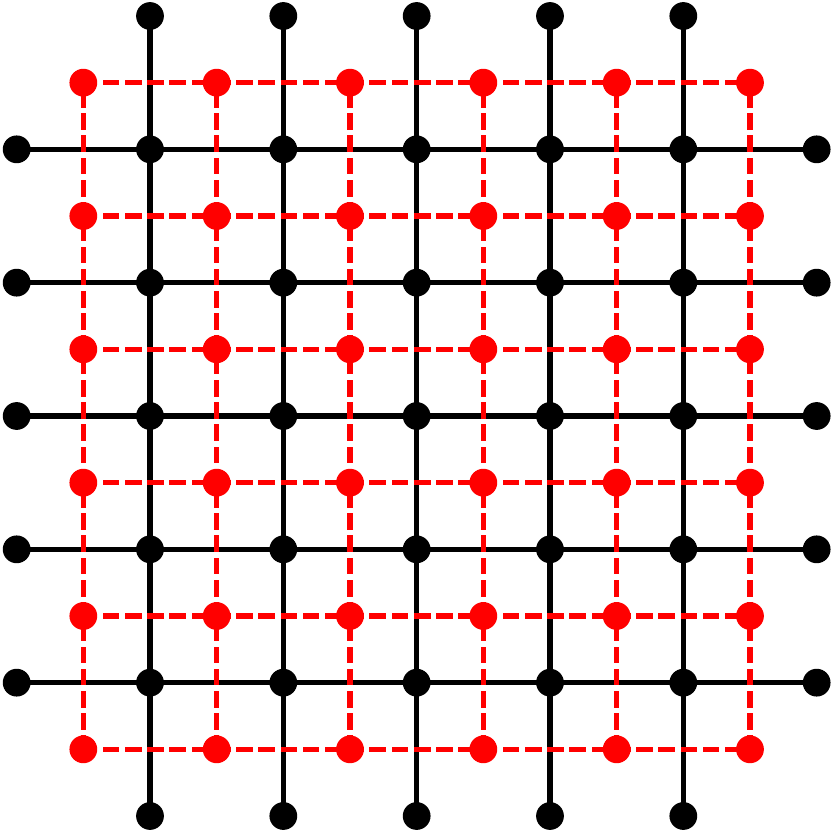}
\caption{The lattice with fixed spin boundary conditions (black) and the dual lattice with free boundary conditions (red).}
\label{fig:latt}
\end{center}
\end{figure}

\begin{figure}[htbp]
\begin{center}
\includegraphics[width=2in]{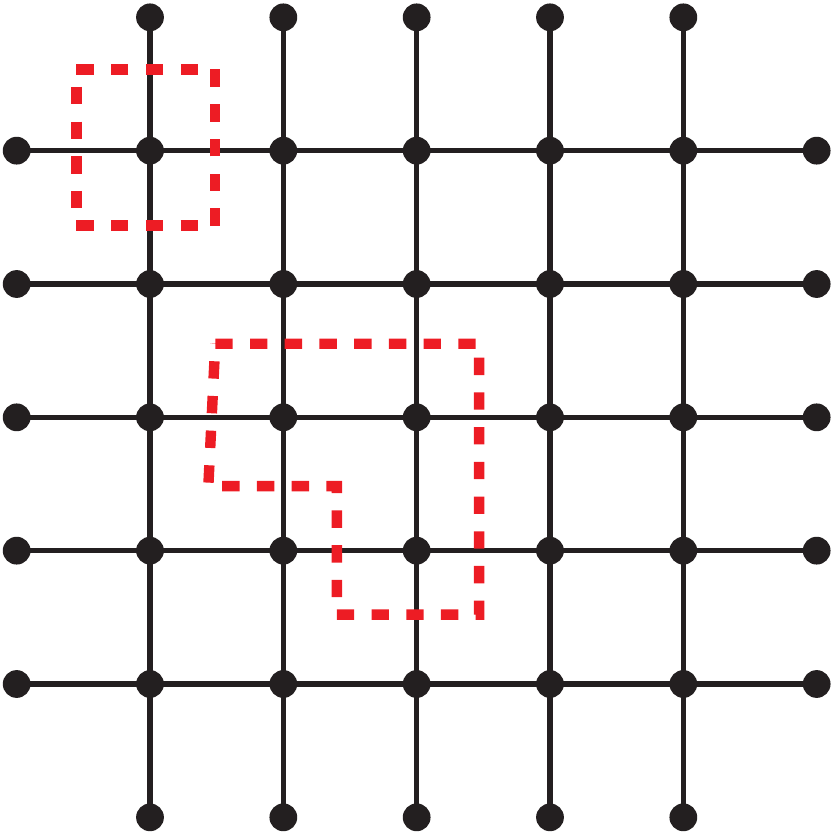}
\caption{Graphical representation of a term in the low temperature expansion of the partition function. The spins on sites inside the regions enclosed by dashed red lines point in a direction opposite to that in which all other spins point.  The spins on the boundary, which are fixed, are not allowed to reverse. }
\label{fig:lt}
\end{center}
\end{figure}

The first demonstration of Kramers-Wannier duality relation between the Casimir force for fixed spin and free boundary conditions \cite{KW,KWp} utilizes a standard argument based on high and low temperature expansions \cite{Kogut}. Figure \ref{fig:latt} shows the Ising lattce subject to fixed boundary conditions along with its dual. The fixed-boundary-condition lattice has bonds represented as solid black lines, and the sites at which the spins sit are shown as black circles.
The expansion that we consider for the partition function of this lattice is the low temperature one, in which we assume a ``ground state'' in which all spins are up, and we expand with respect to down spins. This generates a graphical expansion for the partition function containing terms such as the one displayed in Fig. \ref{fig:lt}. 

\begin{figure}[htbp]
\begin{center}
\includegraphics[width=2in]{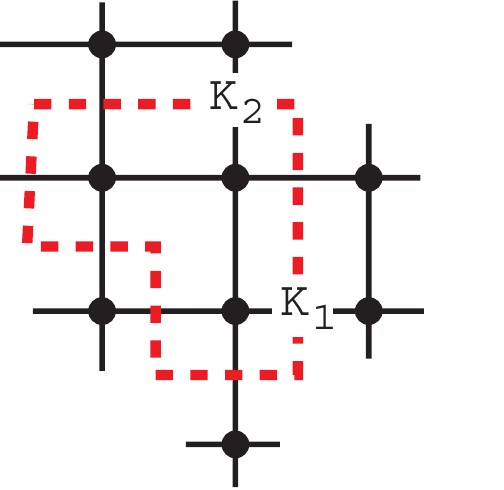}
\caption{A region of reversed spins}
\label{fig:detail}
\end{center}
\end{figure}

Figure \ref{fig:detail} shows one of the sections of reversed spin in \ref{fig:lt}. 
The red dashed lines denote the boundary of that region. As indicated in the figure, the strength of the horizontal bonds is $K_1$ and the strength of the vertical bonds is $K_2$. The contribution to the partition function of a portion of the boundary intersecting a horizontal bond is $e^{-2K_1}$ corresponding to the alteration of the Boltzmann factor associated with the mismatch of a pair of spins; when the portion of the boundary intersects a vertical bond the contribution is $e^{-2K_2}$. 

An important feature of this expansion is that fixed boundary conditions, in which the spins at the periphery of the system are constrained to point up, guarantee that all contributions to the low temperature expansion consist of complete closed boundaries

\begin{figure}[htbp]
\begin{center}
\includegraphics[width=2in]{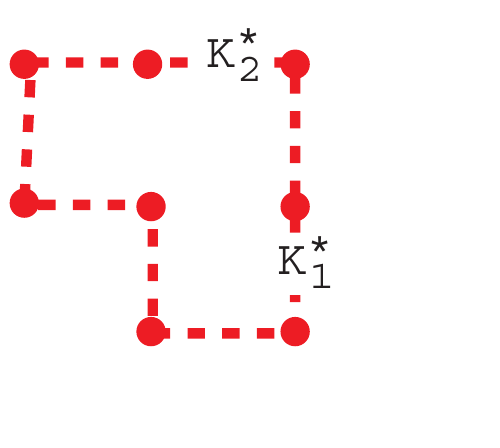}
\caption{The portion of the high temperature expansion corresponding to the detail in the low temperature expansion shown in Fig. \ref{fig:detail}. }
\label{fig:dual}
\end{center}
\end{figure}

The next step in the development is to generate a high temperature expansion for the partition function of the spin system on the dual lattice.  The spins on this lattice interact via nearest neighbor bonds that are both horizontal and vertical as in the original lattice. The strength of the horizontal bonds is $K_1^*$ and of the vertical bonds is $K_2^*$. The high temperature expansion is based on the decomposition 
\begin{eqnarray}
\exp[K\sigma \sigma^{\prime}] & = & \cosh K + \sigma \sigma^{\prime} \sinh K \nonumber \\
& = & \cosh K \left(1 + \sigma \sigma^{\prime} \tanh K \right)
\label{eq:b72}
\end{eqnarray}
The expansion is in the second term in parentheses on the last line of (\ref{eq:b72}). Because there is precisely one such term for each pair of spins and because each spin must appear an even number of times in the expansion, the high temperature expansion is represented graphically by figures like the border shown in Fig. \ref{fig:detail}, this time with the portions of the figure standing for bonds between spins on the dual lattice, as shown in Fig. \ref{fig:dual}. The portion is shown in Fig. \ref{fig:dual}.
Each segment of the portion yields a factor $\tanh K_1^*$ if it is horizontal and $\tanh K_2^*$. Perfect correspondence is achieved if
\begin{eqnarray}
\tanh K_1^* & = & \exp[-2K_2] \label{eq:b73} \\
\tanh K_2^* & = & \exp[-2K_1] \label{eq:b74}
\end{eqnarray}
This is the standard set of duality relations. 

Here, in order that the high temperature expansion also consist of closed contours, we must allow all spins on the dual lattice to vary without restrictions. That is, the boundary conditions on the dual lattice must be free. This establishes the complete duality between the two versions of the two dimensional Ising model.


\begin{thebibliography}{27}
\expandafter\ifx\csname natexlab\endcsname\relax\def\natexlab#1{#1}\fi
\expandafter\ifx\csname bibnamefont\endcsname\relax
  \def\bibnamefont#1{#1}\fi
\expandafter\ifx\csname bibfnamefont\endcsname\relax
  \def\bibfnamefont#1{#1}\fi
\expandafter\ifx\csname citenamefont\endcsname\relax
  \def\citenamefont#1{#1}\fi
\expandafter\ifx\csname url\endcsname\relax
  \def\url#1{\texttt{#1}}\fi
\expandafter\ifx\csname urlprefix\endcsname\relax\def\urlprefix{URL }\fi
\providecommand{\bibinfo}[2]{#2}
\providecommand{\eprint}[2][]{\url{#2}}

\bibitem[{\citenamefont{Fisher and de~Gennes}(1978)}]{fisherdegennes}
\bibinfo{author}{\bibfnamefont{M.~E.} \bibnamefont{Fisher}} \bibnamefont{and}
  \bibinfo{author}{\bibfnamefont{P.~G.} \bibnamefont{de~Gennes}},
  \bibinfo{journal}{Comptes Rendus Hebdomadaires Des Seances De L Academie Des
  Sciences Serie B} \textbf{\bibinfo{volume}{287}}, \bibinfo{pages}{207}
  (\bibinfo{year}{1978}).

\bibitem[{\citenamefont{Fisher and Barber}(1972)}]{fishbarb}
\bibinfo{author}{\bibfnamefont{M.~E.} \bibnamefont{Fisher}} \bibnamefont{and}
  \bibinfo{author}{\bibfnamefont{M.~N.} \bibnamefont{Barber}},
  \bibinfo{journal}{Physical Review Letters} \textbf{\bibinfo{volume}{28}},
  \bibinfo{pages}{1516} (\bibinfo{year}{1972}).

\bibitem[{\citenamefont{Krech}(1997)}]{binary}
\bibinfo{author}{\bibfnamefont{M.}~\bibnamefont{Krech}},
  \bibinfo{journal}{Phys. Rev. E} \textbf{\bibinfo{volume}{56}},
  \bibinfo{pages}{1642} (\bibinfo{year}{1997}).

\bibitem[{\citenamefont{Garcia and Chan}(1999)}]{gc1}
\bibinfo{author}{\bibfnamefont{R.}~\bibnamefont{Garcia}} \bibnamefont{and}
  \bibinfo{author}{\bibfnamefont{M.~H.~W.} \bibnamefont{Chan}},
  \bibinfo{journal}{Physical Review Letters} \textbf{\bibinfo{volume}{83}},
  \bibinfo{pages}{1187} (\bibinfo{year}{1999}).

\bibitem[{\citenamefont{Ganshin et~al.}(2006)\citenamefont{Ganshin,
  Scheidemantel, Garcia, and Chan}}]{gc2}
\bibinfo{author}{\bibfnamefont{A.}~\bibnamefont{Ganshin}},
  \bibinfo{author}{\bibfnamefont{S.}~\bibnamefont{Scheidemantel}},
  \bibinfo{author}{\bibfnamefont{R.}~\bibnamefont{Garcia}}, \bibnamefont{and}
  \bibinfo{author}{\bibfnamefont{M.~H.~W.} \bibnamefont{Chan}},
  \bibinfo{journal}{Physical Review Letters} \textbf{\bibinfo{volume}{97}}
  (\bibinfo{year}{2006}), \bibinfo{note}{075301}.

\bibitem[{\citenamefont{Zandi et~al.}(2004)\citenamefont{Zandi, Rudnick, and
  Kardar}}]{zandi2}
\bibinfo{author}{\bibfnamefont{R.}~\bibnamefont{Zandi}},
  \bibinfo{author}{\bibfnamefont{J.}~\bibnamefont{Rudnick}}, \bibnamefont{and}
  \bibinfo{author}{\bibfnamefont{M.}~\bibnamefont{Kardar}},
  \bibinfo{journal}{Physical Review Letters} \textbf{\bibinfo{volume}{93}},
  \bibinfo{pages}{155302} (\bibinfo{year}{2004}).

\bibitem[{\citenamefont{Dantchev and Krech}(2004)}]{KD}
\bibinfo{author}{\bibfnamefont{D.}~\bibnamefont{Dantchev}} \bibnamefont{and}
  \bibinfo{author}{\bibfnamefont{M.}~\bibnamefont{Krech}},
  \bibinfo{journal}{Physical Review E} \textbf{\bibinfo{volume}{69}}
  (\bibinfo{year}{2004}).

\bibitem[{\citenamefont{Williams}(2004)}]{williams}
\bibinfo{author}{\bibfnamefont{G.~A.} \bibnamefont{Williams}},
  \bibinfo{journal}{Physical Review Letters} \textbf{\bibinfo{volume}{92}}
  (\bibinfo{year}{2004}).

\bibitem[{\citenamefont{Zandi et~al.}(2007)\citenamefont{Zandi, Shackell,
  Rudnick, Kardar, and Chayes}}]{zandi1}
\bibinfo{author}{\bibfnamefont{R.}~\bibnamefont{Zandi}},
  \bibinfo{author}{\bibfnamefont{A.}~\bibnamefont{Shackell}},
  \bibinfo{author}{\bibfnamefont{J.}~\bibnamefont{Rudnick}},
  \bibinfo{author}{\bibfnamefont{M.}~\bibnamefont{Kardar}}, \bibnamefont{and}
  \bibinfo{author}{\bibfnamefont{L.~P.} \bibnamefont{Chayes}},
  \bibinfo{journal}{Physical Review E} \textbf{\bibinfo{volume}{76}}
  (\bibinfo{year}{2007}).

\bibitem[{\citenamefont{Maciolek et~al.}(2007)\citenamefont{Maciolek, Gambassi,
  and Dietrich}}]{maciolek}
\bibinfo{author}{\bibfnamefont{A.}~\bibnamefont{Maciolek}},
  \bibinfo{author}{\bibfnamefont{A.}~\bibnamefont{Gambassi}}, \bibnamefont{and}
  \bibinfo{author}{\bibfnamefont{S.}~\bibnamefont{Dietrich}},
  \bibinfo{journal}{Physical Review E} \textbf{\bibinfo{volume}{76}}
  (\bibinfo{year}{2007}).

\bibitem[{\citenamefont{Hucht}({2007})}]{hucht}
\bibinfo{author}{\bibfnamefont{A.}~\bibnamefont{Hucht}},
  \bibinfo{journal}{{Phys. Rev. Lett.}} \textbf{\bibinfo{volume}{{99}}}
  (\bibinfo{year}{{2007}}), ISSN \bibinfo{issn}{{0031-9007}}.

\bibitem[{\citenamefont{Krech}(1999)}]{krech}
\bibinfo{author}{\bibfnamefont{M.}~\bibnamefont{Krech}},
  \bibinfo{journal}{Journal of Physics-Condensed Matter}
  \textbf{\bibinfo{volume}{11}}, \bibinfo{pages}{R391} (\bibinfo{year}{1999}).

\bibitem[{\citenamefont{Brankov et~al.}(2000)\citenamefont{Brankov, Danchev,
  and Tonchev}}]{danbook}
\bibinfo{author}{\bibfnamefont{I.}~\bibnamefont{Brankov}},
  \bibinfo{author}{\bibfnamefont{D.~M.} \bibnamefont{Danchev}},
  \bibnamefont{and} \bibinfo{author}{\bibfnamefont{N.~S.}
  \bibnamefont{Tonchev}}, \emph{\bibinfo{title}{Theory of critical phenomena in
  finite-size systems : scaling and quantum effects}}
  (\bibinfo{publisher}{World Scientific}, \bibinfo{address}{Singapore ; River
  Edge, NJ}, \bibinfo{year}{2000}).

\bibitem[{\citenamefont{Milton}(2001)}]{miltonbook}
\bibinfo{author}{\bibfnamefont{K.~A.} \bibnamefont{Milton}},
  \emph{\bibinfo{title}{The Casimir effect : physical manifestations of
  zero-point energy}} (\bibinfo{publisher}{World Scientific},
  \bibinfo{address}{Singapore ; River Edge, NJ}, \bibinfo{year}{2001}).

\bibitem[{\citenamefont{Schultz et~al.}(1964)\citenamefont{Schultz, Mattis, and
  Lieb}}]{SML}
\bibinfo{author}{\bibfnamefont{T.~D.} \bibnamefont{Schultz}},
  \bibinfo{author}{\bibfnamefont{D.~C.} \bibnamefont{Mattis}},
  \bibnamefont{and} \bibinfo{author}{\bibfnamefont{E.~H.} \bibnamefont{Lieb}},
  \bibinfo{journal}{Reviews of Modern Physics} \textbf{\bibinfo{volume}{36}},
  \bibinfo{pages}{856} (\bibinfo{year}{1964}).

\bibitem[{\citenamefont{Baxter}(1982)}]{baxter}
\bibinfo{author}{\bibfnamefont{R.~J.} \bibnamefont{Baxter}},
  \emph{\bibinfo{title}{Exactly solved models in statistical mechanics}}
  (\bibinfo{publisher}{Academic Press}, \bibinfo{address}{London ; New York},
  \bibinfo{year}{1982}).

\bibitem[{\citenamefont{Ferdinand and Fisher}(1969)}]{ferdfish}
\bibinfo{author}{\bibfnamefont{A.}~\bibnamefont{Ferdinand}} \bibnamefont{and}
  \bibinfo{author}{\bibfnamefont{M.~E.} \bibnamefont{Fisher}},
  \bibinfo{journal}{Physical Review} \textbf{\bibinfo{volume}{185}},
  \bibinfo{pages}{832} (\bibinfo{year}{1969}).

\bibitem[{\citenamefont{Wu and Hu}(2002)}]{wuandhu}
\bibinfo{author}{\bibfnamefont{M.~C.} \bibnamefont{Wu}} \bibnamefont{and}
  \bibinfo{author}{\bibfnamefont{C.~K.} \bibnamefont{Hu}},
  \bibinfo{journal}{Journal of Physics a-Mathematical and General}
  \textbf{\bibinfo{volume}{35}}, \bibinfo{pages}{5189} (\bibinfo{year}{2002}),
  \bibinfo{note}{times Cited: 8}.

\bibitem[{\citenamefont{Li et~al.}(1991)\citenamefont{Li, Paczuski, Kardar, and
  Huang}}]{li}
\bibinfo{author}{\bibfnamefont{H.}~\bibnamefont{Li}},
  \bibinfo{author}{\bibfnamefont{M.}~\bibnamefont{Paczuski}},
  \bibinfo{author}{\bibfnamefont{M.}~\bibnamefont{Kardar}}, \bibnamefont{and}
  \bibinfo{author}{\bibfnamefont{K.}~\bibnamefont{Huang}},
  \bibinfo{journal}{Physical Review B} \textbf{\bibinfo{volume}{44}},
  \bibinfo{pages}{8274} (\bibinfo{year}{1991}).

\bibitem[{\citenamefont{Evans and Stecki}(1994)}]{es}
\bibinfo{author}{\bibfnamefont{R.}~\bibnamefont{Evans}} \bibnamefont{and}
  \bibinfo{author}{\bibfnamefont{J.}~\bibnamefont{Stecki}},
  \bibinfo{journal}{Physical Review B} \textbf{\bibinfo{volume}{49}},
  \bibinfo{pages}{8842} (\bibinfo{year}{1994}).

\bibitem[{\citenamefont{Abraham}(1978)}]{DBA}
\bibinfo{author}{\bibfnamefont{D.~B.} \bibnamefont{Abraham}},
  \bibinfo{journal}{Communications in Mathematical Physics}
  \textbf{\bibinfo{volume}{60}}, \bibinfo{pages}{181} (\bibinfo{year}{1978}).

\bibitem[{\citenamefont{Jordan and Wigner}(1928)}]{JW}
\bibinfo{author}{\bibfnamefont{P.}~\bibnamefont{Jordan}} \bibnamefont{and}
  \bibinfo{author}{\bibfnamefont{E.}~\bibnamefont{Wigner}},
  \bibinfo{journal}{Z. Physik} \textbf{\bibinfo{volume}{47}},
  \bibinfo{pages}{631} (\bibinfo{year}{1928}).

\bibitem[{\citenamefont{Abraham and Svrakic}(1986)}]{AS1}
\bibinfo{author}{\bibfnamefont{D.~B.} \bibnamefont{Abraham}} \bibnamefont{and}
  \bibinfo{author}{\bibfnamefont{N.~M.} \bibnamefont{Svrakic}},
  \bibinfo{journal}{Physical Review Letters} \textbf{\bibinfo{volume}{56}},
  \bibinfo{pages}{1172} (\bibinfo{year}{1986}).

\bibitem[{\citenamefont{Abraham and Svrakic}(1991)}]{AS2}
\bibinfo{author}{\bibfnamefont{D.~B.} \bibnamefont{Abraham}} \bibnamefont{and}
  \bibinfo{author}{\bibfnamefont{N.~M.} \bibnamefont{Svrakic}},
  \bibinfo{journal}{Journal of Statistical Physics}
  \textbf{\bibinfo{volume}{63}}, \bibinfo{pages}{1077} (\bibinfo{year}{1991}).

\bibitem[{\citenamefont{Kramers and Wannier}(1941{\natexlab{a}})}]{KW}
\bibinfo{author}{\bibfnamefont{H.~A.} \bibnamefont{Kramers}} \bibnamefont{and}
  \bibinfo{author}{\bibfnamefont{G.~H.} \bibnamefont{Wannier}},
  \bibinfo{journal}{Physical Review} \textbf{\bibinfo{volume}{60}},
  \bibinfo{pages}{252} (\bibinfo{year}{1941}{\natexlab{a}}).

\bibitem[{\citenamefont{Kramers and Wannier}(1941{\natexlab{b}})}]{KWp}
\bibinfo{author}{\bibfnamefont{H.~A.} \bibnamefont{Kramers}} \bibnamefont{and}
  \bibinfo{author}{\bibfnamefont{G.~H.} \bibnamefont{Wannier}},
  \bibinfo{journal}{Physical Review} \textbf{\bibinfo{volume}{60}},
  \bibinfo{pages}{263} (\bibinfo{year}{1941}{\natexlab{b}}).

\bibitem[{\citenamefont{Kogut}(1979)}]{Kogut}
\bibinfo{author}{\bibfnamefont{J.~B.} \bibnamefont{Kogut}},
  \bibinfo{journal}{Reviews of Modern Physics} \textbf{\bibinfo{volume}{51}},
  \bibinfo{pages}{659} (\bibinfo{year}{1979}).

\end{thebibliography}
\end{document}